\documentclass[letter]{aa}  

\usepackage{graphicx}
\usepackage{txfonts}
\usepackage{ulem}
\usepackage{threeparttable}
\usepackage[colorlinks=true, linkcolor=blue, citecolor=blue, filecolor=blue, urlcolor=blue]{hyperref}
\begin{document}

   \title{Magnetically Driven Relativistic Jet in the High-Redshift Blazar OH~471}

   \author{
    S. Guo
    \inst{1,2,3}
    \orcid{0000-0003-0181-7656}
\and
    T. An
    \inst{1,2,3}
    \fnmsep\thanks{antao@shao.ac.cn}
    \orcid{0000-0003-4341-0029}
\and
    Y. Liu
    \inst{1}
    \orcid{0000-0001-9321-6000}
\and
    Y. Sotnikova
    \inst{4}
    \orcid{0000-0001-9172-7237}  
    \and
    A. Volvach 
    \inst{5}
    \orcid{0000-0002-3839-3466} 
\and
    T. Mufakharov
    \inst{4,6}
    \orcid{0000-0001-9984-127X}
\and
    L. Chen
    \inst{1}
    \orcid{0000-0002-1908-0536}
\and
    L. Cui
    \inst{7}
    \orcid{0000-0003-0721-5509}
\and
    A. Wang
    \inst{1,2}
    \orcid{0000-0002-7351-5801}
\and
    Z. Xu
    \inst{1}
    \orcid{0000-0003-4853-7619}
\and
    Y. Zhang
    \inst{1}
    \orcid{0000-0001-8256-8887}
\and
    W. Xu
    \inst{2,7}
\and
    Y. A. Kovalev
    \inst{8}
\and
    Y.Y. Kovalev
    \inst{9}
\and
    M. Kharinov
    \inst{10}
\and
    A. Erkenov
    \inst{4}
    \orcid{0000-0002-0321-8588}
\and
    T. Semenova
    \inst{4}
    \orcid{0000-0002-2902-5426}
\and
    L. Volvach
    \inst{5}
    \orcid{0000-0001-6157-003X} 
          }

   \institute{
    Shanghai Astronomical Observatory, Chinese Academy of Sciences (CAS), 80 Nandan Road, Shanghai 200030, China
\and
    School of Astronomy and Space Sciences, University of Chinese Academy of Sciences, No. 19A Yuquan Road, Beijing 100049, China
\and 
    Key Laboratory of Radio Astronomy and Technology, CAS, A20 Datun Road, Beijing, 100101, P. R. China
\and
    Special Astrophysical Observatory of RAS, Nizhny Arkhyz, 369167, Russia
\and
    Radio Astronomy and Geodynamics Department, Crimean Astrophysical Observatory, Katsiveli, RT-22, Crimea      
\and
    Kazan Federal University, 18 Kremlyovskaya St, Kazan 420008, Russia             
\and
    Xinjiang Astronomical Observatory, Chinese Academy of Sciences, 150 Science 1-Street, Urumqi, Xinjiang 830011, China            
\and
    Lebedev Physical Institute (Astro Space Center), Leninsky prospect 53, Moscow, 117997, Russia           
\and
    Max-Planck-Institut f\"ur Radioastronomie, Auf dem H\"ugel 69, Bonn, 53121, Germany
\and
    Institute of Applied Astronomy, Russian Academy of Sciences, Kutuzova Embankment 10, St. Petersburg, 191187, Russia        
             }

   \date{Received March XX, 2024; accepted XX XX, 2024}

 
  \abstract
   {Understanding the mechanisms that launch and shape powerful relativistic jets from supermassive black holes (SMBHs) in high-redshift active galactic nuclei (AGN) is crucial for probing the co-evolution of SMBHs and galaxies over cosmic time.}
   {We study the high-redshift ($z=3.396$) blazar OH~471 to explore the jet launching mechanism in the early Universe.}
   {Using multi-frequency radio monitoring observations and high-resolution Very Long Baseline Interferometry imaging over three decades, we study the milliarcsecond structure and long-term variability of OH~471.}
   {Spectral modelling of the radio flux densities reveals a synchrotron self-absorbed spectrum indicating strong magnetic fields within the compact core. By applying the flux freezing approximation, we estimate the magnetic flux carried by the jet and find that it reaches or exceeds theoretical predictions for jets powered by black hole spin energy via the Blandford-Znajek mechanism. This implies that OH~471 was in a magnetically arrested disk (MAD) state where the magnetic flux accumulated near the horizon regulates the accretion flow, allowing efficient extraction of black hole rotational energy.  }
   {Our study demonstrates the dominance of MAD accretion in powering the prominent radio flares and relativistic jets observed in the radio-loud AGN OH~471 and statistical studies of large samples of high-redshift AGN will shed light on the role of MAD accretion in launching and accelerating the earliest relativistic jets. }


   \maketitle
%

\section{Introduction} \label{sec:intro}

Understanding the jet launching mechanism in high-redshift active galactic nuclei (AGN) is crucial for unravelling the complex interplay between supermassive black hole (SMBH) accretion and feedback processes in the early Universe \citep[e.g.,][]{2005Natur.433..604D, 2008ApJ...676...33D, 2014MNRAS.440.1590D}. The magnetically arrested disk (MAD) model proposes that relativistic jets can be powered efficiently when the magnetic flux threading the black hole reaches a critical threshold \citep{2003PASJ...55L..69N}. Previous studies have analysed AGN jets at redshifts up to 2.37, with most below $z \sim 2$ \citep{2014Natur.510..126Z}, however, confirming or refuting the MAD paradigm in higher redshift blazars can provide unique insights into the earliest epochs of  SMBH accretion physics and jet formation mechanisms.

In this paper, we study the radio source OH~471  (0642+449, J0646+4451), a high-redshift ($z = 3.396$) blazar that serves as a unique laboratory for testing MAD models and probing jet physics in the early Universe.
It was one of the first distant ($z>3$) radio quasars identified in the 1970s \citep{1972AJ.....77..557G,1973Natur.242..394C,1973Natur.243...25C}. 
Early radio observations of OH~471 revealed a complex spectral structure with multiple synchrotron self-absorption (SSA) components on different physical scales peaking between 0.5--20 GHz \citep{1974Natur.249..743G, 1974Natur.250..472J}, indicating emission from several distinct regions in the relativistic jet.
Very Long Baseline Interferometry (VLBI) observations established OH~471 as a blazar with a typical one-sided core-jet morphology on milliarcsecond scales \citep{1980AJ.....85..668M, 1992A&A...260...82G, 1998AJ....115.1295K, 2005AJ....130.2473K, 2013AJ....146..120L,2015AJ....150...58F,2015A&A...583A.100L}.

Multi-frequency VLBI observations reveal a compact core region with low (2\%) fractional polarization and transverse magnetic fields \citep{2015A&A...583A.100L}. This contrasts with the substantially higher fractional polarization in a jet component located $\sim 4$ mas ($\sim$30 parsec) downstream the core, which displays magnetic fields aligning with the jet. 
Furthermore, an extremely high rotation measure (RM) of $9 \times 10^3$ rad $\rm m^{-2}$ (redshift-corrected) is inferred in the inner jet just 0.5 mas from the core \citep{2012AJ....144..105H}. This represents the largest redshift-corrected RM detected among similar high-redshift blazars.

A key motivation of this study is to explore the jet launching mechanism in OH~471 by estimating the magnetic flux and testing MAD models. 
Evidence for sufficient magnetic flux would strongly support magnetically-driven relativistic jets in this high-redshift radio-loud quasar. 

A direct way to estimate the magnetic field strength is to fit the SSA spectrum of OH~471. 
The radio spectrum of OH~471 displays dramatic evolution over decades, with the SSA turnover frequency shifting from $\sim$1 GHz in early observations \citep{1974Natur.249..743G,1974Natur.250..472J} to 10--20 GHz in recent monitoring data \citep{2000KFNTS...3...21V, 2021MNRAS.508.2798S}.  
This remarkable spectral variability likely implies substantial changes in the magnetic field strength, particle densities and opacity close to the SMBH.
However, the spectrum has transitioned to an unusually flat shape between 1--22 GHz since 2015, representing a substantial departure from the earlier SSA-dominated spectrum. Such spectral flattening likely indicates the emergence of a newly generated jet component still embedded in the core.

By connecting changes in the jet structure and kinematics on milli-arcsec (mas) scales to this long-term spectral evolution, we expect to reveal the key structural changes in the inner jet responsible for the spectral change. 
Combining multi-epoch VLBI images over decades \footnote{\url{https://astrogeo.org/} maintained by Leonid Petrov.}
allow us to probe the jet kinematics. 
Identifying newly ejected components associated with the spectral flattening since 2015 could reveal essential clues to the particle acceleration and magnetic field evolution near the central engine.

In this paper, we provide an in-depth analysis of the jet properties and magnetic flux using multi-frequency radio variability modelling and up to three decades of VLBI image data from 1995 to 2022. 
The paper is organized as follows: Section \ref{sec:data} describes the data; Section \ref{sec:results} presents the results; Section \ref{sec:discussion} provides discussion; and Section \ref{sec:summary} summarizes the conclusions. We assume a $\Lambda$CDM cosmology\footnote{Using this cosmological model, 1 mas angular scale corresponds to a projected physical size of 7.635 pc at $z=3.396$, and 1 mas yr$^{-1}=109.5\, c$. } with $\Omega_{\rm m} = 0.27$, $\Omega_{\rm \Lambda} = 0.73$, $H_0 = 70$~km\,s$^{-1}$ Mpc$^{-1}$. 

\section{Observations and Data Analysis}\label{sec:data}

This study utilizes two main radio datasets to investigate the spectral evolution and jet structure of radio OH~471.

The first dataset is comprised of radio flux density light curves monitored by the RATAN-600 radio telescope at the Special Astrophysical Observatory of Russian Academy of Sciences \citep{1993IAPM...35....7P}. Continuous monitoring of OH~471 has been carried out since 1997 at six frequencies: 1.1, 2.3, 4.7, 7.7/8.2, 11.2, and 21.7/22.3 GHz \citep{2014A&A...572A..59M,2017AN....338..700M}. The mean time interval is approximately 65 days.
Five epochs of data points observed by the RT-32 Zelenchukskaya and Badary radio telescopes \citep{2019..VLBI..Quasar} at 5.05 and 8.63 GHz are also included.
In total, the RATAN-600 and single-dish light curves (Figure \ref{fig-vlbi-and-ratan}) include over 105 epochs spanning 15 years from 2009 to 2024 (Appendix \ref{app:ratan}). This long-term multi-frequency monitoring provides excellent temporal sampling to quantify the evolution of the radio spectrum.

The second dataset consists of VLBI data at 2.3, 5, 8.4, and 15 GHz observed between 1995 and 2022. The 2.3, 5 and 8.4 GHz VLBI data are primarily obtained from the geodetic observations archived in the Astrogeo database \citep[e.g.][]{2002ApJS..141...13B}. The 15.4 GHz VLBI data are from the Monitoring Of Jets in Active galactic nuclei with VLBA Experiments (MOJAVE) programme \footnote{\url{https://www.cv.nrao.edu/MOJAVE/}} \citep{2018ApJS..234...12L}. 
We used calibrated visibility data from these databases that had undergone initial calibration steps. We only performed a few iterations of self-calibration and made mode fitting to study the change of the jet structure over time. 
The details of VLBI data processing are described in Appendix \ref{app:vlbi}.

Combining multi-frequency light curves, VLBI imaging, and spectral modelling enables a robust investigation connecting the structural and spectral evolution of OH~471.

\section{Results}\label{sec:results}

\subsection{Radio variability}\label{sec:varability}

Figure \ref{fig-vlbi-and-ratan} presents the flux density variations of OH~471 observed by VLBI (top panel) and RATAN-600 (bottom panel) over a wide range of radio frequencies. 
Two major flares are evident --  around March 2003 and October 2008 -- visible only 
at the higher frequencies of 8 and 15 GHz. The higher-frequency ($\geq4.7$ GHz) flux densities gradually decrease after 2009, likely indicating a period of reduced activity. In contrast, the 2-GHz flux densities show a slow rise from 1997, peaks around December 2013, and declines thereafter. This behaviour likely reflects the shift in the SSA turnover frequency over time and the inherent lag in the low-frequency emission response to core flare activity.

The RATAN-600 lightcurves start in October 2009, although the initial onset of the 2008 flare is not captured due to limited temporal coverage. Nevertheless, the decline from the peak flux density is well documented. Following this flare, the light curves exhibit significant complexity and irregularity, suggesting possible overlapping events or a highly variable source structure. The 2008 flare declined until 2020,  after which the flux densities showed a pronounced rise, particularly at the higher observed frequencies. This rise
suggests the emergence of a new jet component.
Continuous long-term monitoring is 
desirable to fully capture the evolution of this late-arising event, as it may analogously trace 
activities similar to that observed during the earlier prominent flares in the 2000s. Studying these prominent flares can help understand the possible connection between the flaring activity and the evolution of the jet structure in high-redshift blazars like OH~471.

\subsection{Jet structure and Kinematics}\label{sec:jetkinematics}

Previous studies have revealed a jet extending 10 
mas east of the core \citep[][]{1992A&A...260...82G} and up to 30 mas at 1.4 GHz (MOJAVE webpage)
\footnote{\url{https://www.cv.nrao.edu/MOJAVE/sourcepages/0642+449.shtml}}, while higher frequency ($\nu \geq 2.3$ GHz) VLBI resolves more compact structure within $\sim$2 mas \citep{1980AJ.....85..668M, 1998AJ....115.1295K, 2005AJ....130.2473K, 2013AJ....146..120L, 2015AJ....150...58F}.
Multi-epoch VLBI observations have revealed apparent superluminal speeds up to $8.5\, c$ and constrained key jet parameters including the Lorentz factor $\Gamma = 5.4$, viewing angle $\theta_{\rm j} = 0.8\degr$, and intrinsic opening angle of $\phi_{\rm j} = 0.3\degr$ \citep{2009A&A...507L..33P, 2013AJ....146..120L}. These studies revealed both fast-moving and stationary jet features, indicating the coexistence of propagating shocks and recollimation sites or jet-ISM interactions.

We analyzed over 50 epochs of 8.4-GHz and 2.3-GHz archival VLBI data spanning 1996--2022 from the Astrogeo database to qualify jet motions. The high resolution and long-time baseline provided by these observations enable robust constraints on jet proper motions. 
We measure a proper motion of $\mu=-0.0025 \pm 0.0085$ mas yr$^{-1}$ for component J1 located $\sim$3.5 mas from the core, and $0.040 \pm 0.003$ mas yr$^{-1}$ (corresponding to $4.4 \pm 0.3 \,c$) for J2 at 0.5--1 mas from the core (Figure \ref{fig-propermotion}). 
The stationary jet component J1 shows an extended morphology, and the polarized flux density along the southern edge of this feature 
could suggest either an interaction between the jet and the ambient medium \citep{2015A&A...583A.100L} or the presence of a helical magnetic field component roughly aligned with the jet.
As the 8-GHz resolution differs from 15-GHz MOJAVE data, we do not directly compare our proper motion of J2 to literature values \citep{2013AJ....146..120L}, which probe distinct jet structures.

The close temporal association between the emergence of jet components and radio flares supports a direct connection between the flaring activity and the ejection of a new jet component from the central engine.
We found that J2 was ejected earlier than the monitored lightcurves (i.e. before 1996), based on backward extrapolation. The observed variability combines the ensemble core and jet emission. If each flare produces a component, but the VLBI resolution cannot distinguish them, the detected motion represents overlapping components. This could explain why individual jet features do not always correspond to single flares. Alternatively, J2 may have decelerated during its outward motion, causing an underestimation of its initial speed. Dividing the J2 data into pre- and post-2016 segments reveals faster motion ($0.053 \pm 0.002$ mas yr$^{-1}$) initially. Back-extrapolating this speed implies an ejection time around 2001, roughly consistent with the strongest observed flare during the past three decades.

\subsection{Radio spectra} \label{sec:spectra}

Long-term radio spectral monitoring of OH~471 has revealed significant variability \citep[e.g.,][]{2003IAUJD..18E...8V}. 
In the 1970s, the spectrum showed a synchrotron self-absorption turnover around 1 GHz and a flat spectrum at higher frequencies, possibly due to a combination of multiple SSA components  \citep{1974Natur.249..743G,1974Natur.250..472J}. However, from the 1980s onward, this low-frequency SSA component faded away, while a new high-frequency SSA component emerged and became dominant.
Since the 1990s, regular (1-22) GHz observations by RATAN-600 revealed strong turnover frequency evolution over time \citep{2012A&A...544A..25M,2021MNRAS.508.2798S}. The turnover was around 20 GHz in May 1998 \citep{2000KFNTS...3...21V} and around 15 GHz in November 1998  \citep{2000A&A...363..887D}, dropping to 11.2 GHz by July 2002 \citep{2005A&A...432...31T}. 
Since 2012, it has remained variable around 17 GHz  \citep{2021MNRAS.508.2798S}.

These long-term observations have revealed clear variability in both the flux density and peak frequency of the inverted spectrum of OH~471 over decades.
To extract physical insights, we modelled each radio spectrum using a synchrotron self-absorption model (Appendix \ref{app:ssa}), fitting for the turnover frequency $\nu_{\rm m}$, flux density at the turnover $S_{\rm m}$, the optically thick spectral index $\alpha_{\rm thick}$, and the optically thin spectral index $\alpha_{\rm thin}$ (see a typical SSA spectrum in Figure \ref{fig:SSA}).
    
Figure \ref{fig:SSA} shows the remarkable evolution of the radio spectrum from 2009 to 2024 based on multi-frequency monitoring data. The observations suggest at least two distinct SSA components: a low-frequency component (turnover $\lesssim 5$ GHz) corresponding to the steady-state radio core, and a high-frequency flaring component (turnover $>5$ GHz). This is consistent with conclusions from early studies \citep{1974Natur.249..743G}.  However, our data only allows fitting a single SSA model. The high-frequency component evolves faster and dominates the overall spectral changes. The new jet component associated with the 2003 and/or 2008 flares gradually faded over 2009--2019, causing $\nu_{\rm m}$ and $\alpha_{\mathrm{thick}}$ to decrease, until another new high-frequency component emerged after 2020, shifting the turnover back to higher frequencies. 
The joint decrease of $\nu_{\rm m}$ and $S_{\rm m}$ over 2009--2019 may imply a steady decline in the magnetic field strength and emitting particle densities in the core region. This is most straightforwardly explained by a reduction in the jet power from the central engine during this period. The combined changes in $\alpha_{\mathrm{thick}}$, $\nu_{\rm m}$ and $S_{\rm m}$ support the hypothesis of a major reconfiguration of the jet structure on sub-pc scales, with the formation of a new magnetized component. Currently, the flux densities of OH~471 are still in a rising phase. Closely monitoring this new component will help understand the details of jet production and evolution.

Our approach of fitting a single SSA spectrum to the integrated radio flux densities has inherent limitations, as the observed flux contains contributions from both the core and jet components. Separating the core and jet reliably requires broader frequency coverage from VLBI imaging. By fitting a single spectrum to the limited frequency data points, there are uncertainties introduced into the derived parameters like the magnetic field strength, though the core dominates the overall flux density in all epochs. We have attempted to fit the radio spectrum using two SSA components (i.e., a radio core and an inner jet), with the core being the dominant SSA component at high frequencies. Comparing the results with those from a single SSA component, we found that the flux density of the core decreases and the peak frequency increases. The combined effect of these changes leads to an increase in the revised $B_{\rm SSA}$, further supporting the paper's conclusion that the jet magnetic flux in OH~471 is higher than in the MAD prediction. However, due to the current sparse number of observational frequency points, we cannot accurately constrain the SSA parameters. Although using the integrated flux density and a single SSA component to fit the radio spectrum is not perfect, it provides a way to constrain the source properties and magnetic fields in the vicinity of the black hole. These limitations should be kept in mind when interpreting the results of this study. Future multi-frequency VLBI studies can overcome this limitation by robustly separating the core and jet contributions for more reliable spectral modelling.
 
\subsection{Brightness temperature of the VLBI core}

OH~471 stands out as one of the most luminous and beamed high-redshift blazars \citep[e.g.,][]{2011MNRAS.415.3049O}. High brightness temperatures exceeding $10^{12}$ K observed for the OH~471 core (Table \ref{tab:parameters}) are consistent with the expectations for a highly Doppler-boosted relativistic jet pointed roughly toward our line of sight (Section \ref{sec:jetkinematics}). 
We also find that the maximum brightness temperature $T_{\rm b,max}$ shows a systematic decreasing trend with increasing observation frequency:
\begin{itemize} \setlength\itemsep{0em}
    \item $T_{\rm b} > 4 \times 10^{13}$ K at 1.6 GHz from RadioAstron space VLBI observations \citep{2015A&A...583A.100L}; 
    \item $T_{\rm b} \sim 3.3 \times 10^{12}$ K at 5 GHz from the VLBI Space Observatory Program observations \citep{2008ApJS..175..314D};
    \item $T_{\rm b} = 1.2-4.3 \times 10^{12}$ K at 15 GHz from ground-based VLBA observations \citep{2005AJ....130.2473K}; 
    \item $T_{\rm b} = 1.1-1.6 \times 10^{11}$ K at 86 GHz from the Global mm-VLBI Array \citep{2000A&A...364..391L,2008AJ....136..159L}.
\end{itemize}

The decreasing trend in $T_{\rm b,max}$
of the OH~471 core with increasing observation frequency
can be explained by a combination of effects: synchrotron self-absorption causes the core to become more optically thin and extended at higher frequencies; the SSA turnover around 8 GHz (Section \ref{sec:varability}) suggests that emissions at frequencies below this peak are in the optically thick regime, contributing to higher observed temperatures; 
higher resolution available at higher frequencies may resolve substructure and gradients in the core magnetic field, particle densities, and relativistic beaming; different emission mechanisms can dominate across frequencies; unresolved sub-components likely contribute to the complex spectral behaviour.

\section{Discussion}\label{sec:discussion}

As one of the most distant blazars known, OH~471 provides a unique window into jet production and particle acceleration in the vicinity of an actively growing SMBH in the early Universe. Studying the variability, structural changes and magnetic field evolution of OH~471 is key to unravelling the physical processes driving powerful high-redshift jets.

The magnetic flux paradigm proposes that jet power is fundamentally determined by the accumulation of poloidal magnetic flux around the black hole \citep{2013ApJ...764L..24S}. Poloidal flux refers to the vertical magnetic field component that threads both the black hole and the inner accretion disk. 
In MADs, strong poloidal fields obstruct the inner accretion flow, choking the accretion and creating a highly magnetized inner region \citep{2011MNRAS.418L..79T}.
In  MAD states, the strong vertical magnetic fields extract the spin energy of the black hole through the Blandford-Znajek process \citep{1977MNRAS.179..433B}, 
generating jets with kinetic powers comparable to or exceeding the accretion power \citep{2003PASJ...55L..69N}.
Without adequate magnetic flux accumulation, AGN cannot achieve MAD conditions, resulting in weaker jets compared to accretion power.

Observations support links between the magnetic flux and the jet power: e.g., MAD jets exhibit more ordered inner disk fields and higher core rotation measures \citep{2014Natur.510..126Z}; intermittent jet activity implies fluctuating magnetic flux levels \citep{2013ApJ...764L..24S}; jet power correlates with radio core luminosity \citep{2015MNRAS.449..316N}. These manifest that creating powerful jets relies on accumulating strong poloidal magnetic flux, beyond just black hole spin. 

We can test if OH~471 has achieved the MAD state by comparing its estimated jet magnetic flux to the predicted MAD flux, thus assessing the role of magnetic flux in powering the powerful high-redshift jet in OH~471.

Models of jet launching propose that relativistic jets can be efficiently powered once the poloidal flux surrounding the black hole exceeds a critical threshold via the Blandford-Znajek mechanism. 
Simulations indicate that the MAD state occurs when magnetic flux reaches $\sim50\left(\dot{M} r_{\mathrm{g}}^2c\right)^{1/2}$ \citep{2011MNRAS.418L..79T}, in which $\dot{M}$ is the accretion rate, $r_\mathrm{g}$ the gravitational radius, and $c$ the speed of light. 
The mass accretion rate is $\dot{M} = L_{\mathrm{acc}} / (\eta c^2)$, where $L_{\mathrm{acc}}$ is the accretion power, $\eta$ the efficiency (typically 0.1--0.4). The black hole mass ($M_{\mathrm{BH}}$) sets the gravitational radius via $r_\mathrm{g} = G M_{\mathrm{BH}}/c^2$, where $G$ is the gravitational constant. Combining these quantities, the predicted MAD magnetic flux is $\Phi_{\mathrm{MAD}} \approx 50\sqrt{ L_{\mathrm{acc}} r_\mathrm{g}^2/\eta c} = 2.4 \times 10^{25} [\frac{\eta}{0.4}]^{-1/2} [\frac{M_{\rm BH}}{M_\odot}][\frac{L_{\rm acc} }{1.26 \times 10^{47} \mathrm{erg \, s}^{-1}}]^{1/2} [\mathrm{G \, cm^2}]$.
The accretion luminosity \( L_{\mathrm{acc}} \) is calculated using the bolometric luminosity \( L_{\mathrm{bol}} \) and an accretion efficiency, giving \( L_{\mathrm{acc}} = \frac{L_{\mathrm{bol}}}{\eta} \). The bolometric luminosity was estimated from the optical C IV line luminosity.
OH~471 presents strong optical emissions, which can be used to estimate
$L_{\mathrm{acc}}\approx10L_{\mathrm{BLR}}\approx88.2L_{\mathrm{C\ IV}}\approx6.5\times10^{46}$ erg s$^{-1}$ \citep{2006ApJ...637..669L, 1991ApJ...373..465F, 2008MNRAS.387.1669G}.
With $\eta\approx0.4$ \citep[][]{2014Natur.510..126Z}, the estimated MAD flux is $\Phi_{\rm MAD}\approx 1.7 \times 10^{25} [\frac{M_{\rm BH}}{M_{\odot}}]$ [G cm$^{2}$].

While we lack direct observations of the jet magnetic flux in OH~471, we can estimate it from synchrotron self-absorption magnetic field strength ($B_{\rm SSA}$) derived from the spectrum fitting and core size measurement. 
We estimate the SSA magnetic field strength 
following \citet{1983ApJ...264..296M} with details presented in Appendix \ref{app:mag}. $B_{\rm SSA}$ depends on turnover frequency, peak flux density, core size, and the Lorentz factor (Appendix \ref{app:ssa}). The core size is derived from VLBI model fitting (Appendix \ref{app:vlbi}). Following \citet{2021A&A...652A..14C}, $\Phi_{\rm jet} = 0.8 \times 10^{25} f(a*) (1+\sigma)^{1/2} [\frac{M_{\rm BH}}{M_{\odot}}] B$ [G cm$^{2}$], where $\sigma$ is the jet magnetisation parameter, $f(a*)$ is a dimensionless parameter related to the black hole spin, $B$ is the magnetic field strength at 1 parsec from the black hole.
$f(a*) = [1 + \sqrt{1 - a^2}]/a$ is a function of the black hole spin $a$.  For a rapidly rotating black hole, which is expected for powerful jet sources such as blazars, \( f(a*) \approx 1 \) (Appendix \ref{eq:PhiMAD}).
The magnetic field strength at a distance of 1 pc from the black hole, $B_{\rm 1pc}$, was estimated using the relation $B_{\rm 1pc} = B_{\rm core} \times (h_{\rm core} / {\rm 1 pc})^{-\beta}$, where $h_{\rm core}$ is the distance between the VLBI core and the central engine, $B_{\rm core}$ is the core magnetic field strength derived from the SSA fitting, and $\beta$ is the power-law index describing the magnetic field decay with distance, as expected for a conical jet geometry \citep{1979ApL....20...15B}. In this study, we adopt $\beta = 1$, which corresponds to the magnetic field being inversely proportional to the distance from the black hole. 
Given the conical jet geometry, the angular size of a spherical homogeneous synchrotron source at the turnover frequency \( \theta(\nu_\mathrm{m}) \) is approximately equal to the angular size observed at another frequency \( \theta(\nu) \) multiplied by the ratio of the frequencies. 
Further details on the calculation of $B_{\rm 1pc}$ can be found in Appendix \ref{app:mag}.

To independently estimate the magnetic field strength in the VLBI core region, we performed a joint analysis of the simultaneous 8.4 and 15.4 GHz images on 2011 March 5 to determine the core shift. We found a typical core shift of ($0.09 \pm 0.01$) mas between these frequencies. Using the method given by \citet{1998A&A...330...79L}, assuming equipartition between the particle and magnetic field energy densities, and assuming a viewing angle of $0.25\degr$, we estimate a magnetic field strength at a distance of 1 pc from the black hole of $B_{\rm 1pc} = (4.2 \pm 0.8)$ G. This is slightly lower than the value of ($10 \pm 2$) G derived from the SSA fitting, but the two values agree to within $3\sigma$. The core shift estimate thus provides an independent confirmation of the high magnetic field strength in the OH 471 jet, although both methods rely on several simplifying assumptions, so the true uncertainties are likely to be larger than the formal errors reported here.  In addition, we have only one epoch of quasi-simultaneous data, so the core-shift calculation here represents a rough assessment of the magnetic field strength derived from the SSA spectrum fit. Higher-resolution multi-frequency VLBI observations are desirable to refine these magnetic field estimates and to probe the magnetic field structure of the jet in more detail.

The calculated $B_\mathrm{SSA}$, $B_\mathrm{1pc}$, and magnetic flux $\Phi_\mathrm{jet}$ are shown in Figure 4. $B_\mathrm{SSA}$ does not exhibit systematic variations over time. 
$B_\mathrm{SSA}$ depends on multiple parameters ($B_\mathrm{SSA} \propto \nu_\mathrm{m}^5 \theta^4 S_\mathrm{m}^{-2} \delta$) that vary differently over time. These combined effects essentially cancel out, leading to a relatively stable $B_\mathrm{SSA}$. Physically, $B_\mathrm{SSA}$ represents the magnetic field strength right at the SSA photosphere. This photosphere likely moves inward/outward as the jet conditions evolve, sampling different field strengths across a range of radii.
In contrast, $B_\mathrm{1pc}$ approximates the field strength at a fixed 1 parsec distance. This single location maintains a higher field when the jet is actively flaring with fresh particle acceleration. As the flare fades over time, the field decreases at that fixed point.
As $B_\mathrm{1pc}$ is more heavily dependent on $\nu_\mathrm{m}$ through the $\nu_\mathrm{m}^5$ scaling, $B_\mathrm{1pc}$ follows a similar evolution trend with $\nu_\mathrm{m}$.

The inferred $\Phi_\mathrm{jet}$ exceeds the predicted $\Phi_\mathrm{MAD}$ in 2009--2012, when the source was in a declining phase after the large 2008 flare and the newly generated jet had not yet dimmed. From 2015--2020, $\Phi_\mathrm{jet} \approx \Phi_\mathrm{MAD}$, indicating magnetic flux accumulation between flares. The period of 2021 December -- 2024 February represents the rising phase of a new flare whose peak is yet to be determined, and in this period $\Phi_\mathrm{jet}$ is significantly higher than $\Phi_\mathrm{MAD}$, indicating that the magnetic flux has now exceeded the critical threshold needed to trigger a new jet ejection episode. The strong magnetic flux explains how OH~471 is able to launch such a powerful jet. 

The temporal evolution of $\Phi_\mathrm{jet}$ relative to $\Phi_\mathrm{MAD}$ implies a cyclical flux accumulation process, where sufficient magnetic flux builds up to launch a new jet component, followed by a declining phase as the new jet expands and fades. Monitoring the magnetic flux over multiple flaring cycles can further test this magnetic paradigm for powering recurring jets.

To maintain the MAD state for long periods of time, significant magnetic flux is likely to have accumulated early on near the event horizon of the black hole.
This type of accretion is thought to fuel the rapid growth of SMBHs at high redshifts. This scenario suggests a link between the jet's current activity and the AGN's early accretion history. The large reservoir of magnetic flux from past accretion builds up over timescales much longer than the jet variability, enabling the AGN to periodically reach MAD conditions and drive powerful jets intermittently. Despite fluctuations in magnetic flux, the stored flux near the black hole ensures continued jet activity following accretion episodes.

In Figure \ref{fig-relation}, we plot the measured jet magnetic flux $\Phi_{\rm jet}$ versus $L_{\rm acc}^{1/2} M_{\rm BH}$ for OH~471 compared to the sample from \citet{2014Natur.510..126Z}. Their sample consists of 76 radio-loud AGN, including 68 blazars and 8 nearby radio galaxies. \citet{2014Natur.510..126Z} used published measurements of the core-shift effect to calculate the jet magnetic flux for each source. Since $\Phi_{\rm jet}$ varies with time and $M_{\rm BH}$ is not determined for OH~471, we show the range that spans the $\Phi_{\rm jet}$ variations and $M_{\rm BH} = 10^{8-10} M_\odot$.
The (\(\Phi_{\rm jet}\)) value derived for OH~471 broadly aligns with the previously established relationship between jet magnetic flux and the combined quantity (\(L_{\rm acc}^{1/2} M_{\rm BH}\))  for powerful jets from $z = (0.07 - 2.37)$. While our calculation involves uncertainties discussed before in this section, the overall consistency with lower-redshift AGN jets suggests the magnetic flux paradigm remains applicable to powerful jets at  higher redshifts. The agreement between OH~471 and the relation for lower-$z$ sources hints at the potential universality of magnetically-driven launching processes of powerful jets from the local Universe to earlier cosmic epochs.     
Further studies of additional high-redshift jetted AGN can solidify the connection between magnetic flux and jet power from the early AGN population to their modern counterparts.

The magnetic field estimates presented in this paper are based on the SSA spectrum fitting of the entire source. However, this approach has some limitations due to the mixing of the core and inner jet components at lower frequencies (e.g. 2.3 and 5 GHz), which can introduce uncertainties in the derived magnetic field strengths. To obtain more accurate magnetic field estimates and to probe the magnetic field structure of the jet in detail, high-resolution multi-frequency VLBI observations are needed. These observations will allow us to more precisely determine the core shift between different frequencies, providing an independent constraint on the magnetic field strength. By comparing the magnetic field estimates derived from the core shift analysis with those obtained from the SSA spectrum fitting, we will be able to assess the robustness of our results and better constrain the uncertainties. Furthermore, the high-resolution images will allow us to study the evolution of the magnetic field along the jet, providing insights into the collimation and acceleration processes of the jet.

\section{Conclusion and Summary}\label{sec:summary}

As a radio-loud quasar at $z>3$, OH~471 provides a unique window into jet formation and evolution in the early Universe. It shows multiple observational signatures that are consistent with the predictions of a magnetically arrested disk based on the magnetic flux paradigm for powering relativistic jets -- a system in which strong magnetic flux accumulation enables efficient extraction of black hole spin energy to power relativistic jets. 
These indications include: extremely high brightness temperatures exceeding $10^{13}$ K measured with VLBI on parsec scales \citep{2000A&A...364..391L}, implying efficient jet formation from the core region; 
large rotation measures detected in the inner jet \citep{2012AJ....144..105H, 2011MNRAS.415.3049O}, which could arise from ordered magnetic field components and/or dense magnetised plasma screens local to the jet formation region;
long-term variability in the radio flux density and spectral shape \citep[][and the present paper]{2019ApJ...874...43L, 2021MNRAS.508.2798S},
consistent with the recurrent ejection of new synchrotron self-absorbed components seen in many AGN jets including OH~471, which potentially arise from episodic accumulation and release of magnetic flux driving transient MAD episodes near the black hole; 
and an estimated jet magnetic flux that surpasses theoretical predictions for the critical MAD threshold needed to power a relativistic jet (the present paper). All these observations support a scenario where cyclic magnetic flux accumulation likely enables this high-redshift blazar to periodically achieve magnetically arrested states that give rise to energetic jet ejections from the vicinity of the central SMBH. 

This finding provides important insights into existing models of the earliest growth of supermassive black holes. Powerful flares and extremely relativistic jets observed in high-redshift AGN likely require achieving MAD conditions early, allowing time to accumulate magnetic flux. We expect other high-$z$ jetted AGN to follow this magnetic paradigm, with the most powerful systems already reaching MAD levels. Studying a large statistical sample for correlations between jet power, accretion rate, and magnetic flux can firmly establish the importance of magnetic flux accumulation in generating the earliest AGN jets. Future multi-band observations are crucial to understanding if the jet dynamics in OH~471 differ from similar lower-$z$ objects, potentially revealing changes in jet production and collimation over cosmic time. 

While the MAD model provides rich qualitative predictions for the accretion and jet dynamics around black holes, it does not provide universal quantitative relations directly linking magnetic flux accumulation and jet cycles to system parameters. The complex interplay between black hole mass, accretion rate, disk properties, and GRMHD processes makes accurate quantitative modelling a challenge. Nevertheless, the MAD framework provides important insights into the physical mechanisms likely to drive the episodic powerful jets observed from sources such as OH~471. Despite the limitations of quantitative modelling of individual objects, multi-wavelength observations may help to distinguish MAD signatures.

\begin{acknowledgements}
This research is partly supported by the National SKA Program of China (2022SKA0120102, 2022SKA0130103). 
SGG is supported by the Youth Innovation Promotion Association CAS Program under NO 2021258. 
TA and ZJX are supported by the FAST Special Program (NSFC 12041301).
YQL is supported by the Shanghai Post-doctoral Excellence Program and Shanghai Sailing Program (grant number 23YF1455700) and China Postdoctoral Science Foundation (certification number: 2023M733625). 
ALW is thankful for the financial support received from the University of Chinese Academy of Sciences and appreciates the support and hospitality provided by the SKA Observatory and Jodrell Bank Centre for Astrophysics at the University of Manchester.
YYK was supported by the M2FINDERS project which has received funding from the European Research Council (ERC) under the European Union’s Horizon2020 Research and Innovation Programme (grant agreement No 101018682). 
The RATAN-600 observations were supported by the Ministry of Science and Higher Education of the Russian Federation under the contract 075-15-2024-541.
We are grateful to N.A. Nizhelsky, G.V. Zhekanis, and P.G. Tsybulev (Special Astrophysical Observatory of RAS, Russia).
We would like to extend our sincere gratitude to Eduardo Ros for his insightful comments and suggestions during the review of our manuscript, which were invaluable for the improvement of this work.
This work used resources from the China SKA Regional Centre prototype.
We acknowledge the use of archival calibrated VLBI data from the Astrogeo Center data base maintained by Leonid Petrov.
This research has made use of data from the MOJAVE database that is maintained by the MOJAVE team \citep{2018ApJS..234...12L}.
The National Radio Astronomy Observatory is a facility of the National Science Foundation operated under cooperative agreement by Associated Universities, Inc.
\end{acknowledgements}

\clearpage

\begin{figure*}
\centering
\includegraphics[width=0.95\textwidth]{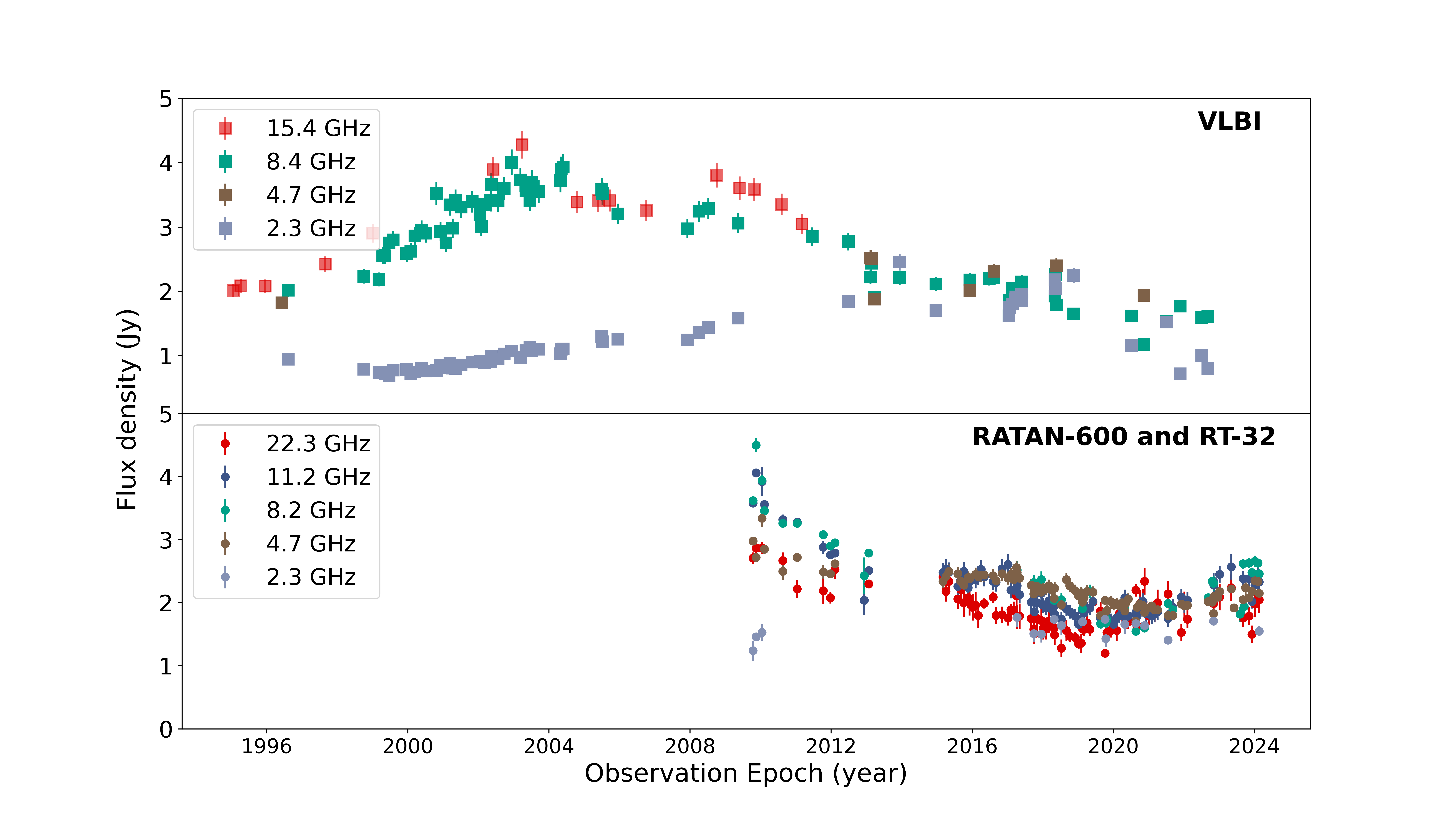}
\caption{Radio light curves of OH~471 made with the VLBI (top panel) , RATAN-600 and RT-32 (bottom panel) observations. 
\label{fig-vlbi-and-ratan}}
\end{figure*}

\begin{figure*}
\centering
\includegraphics[width=0.48\textwidth]{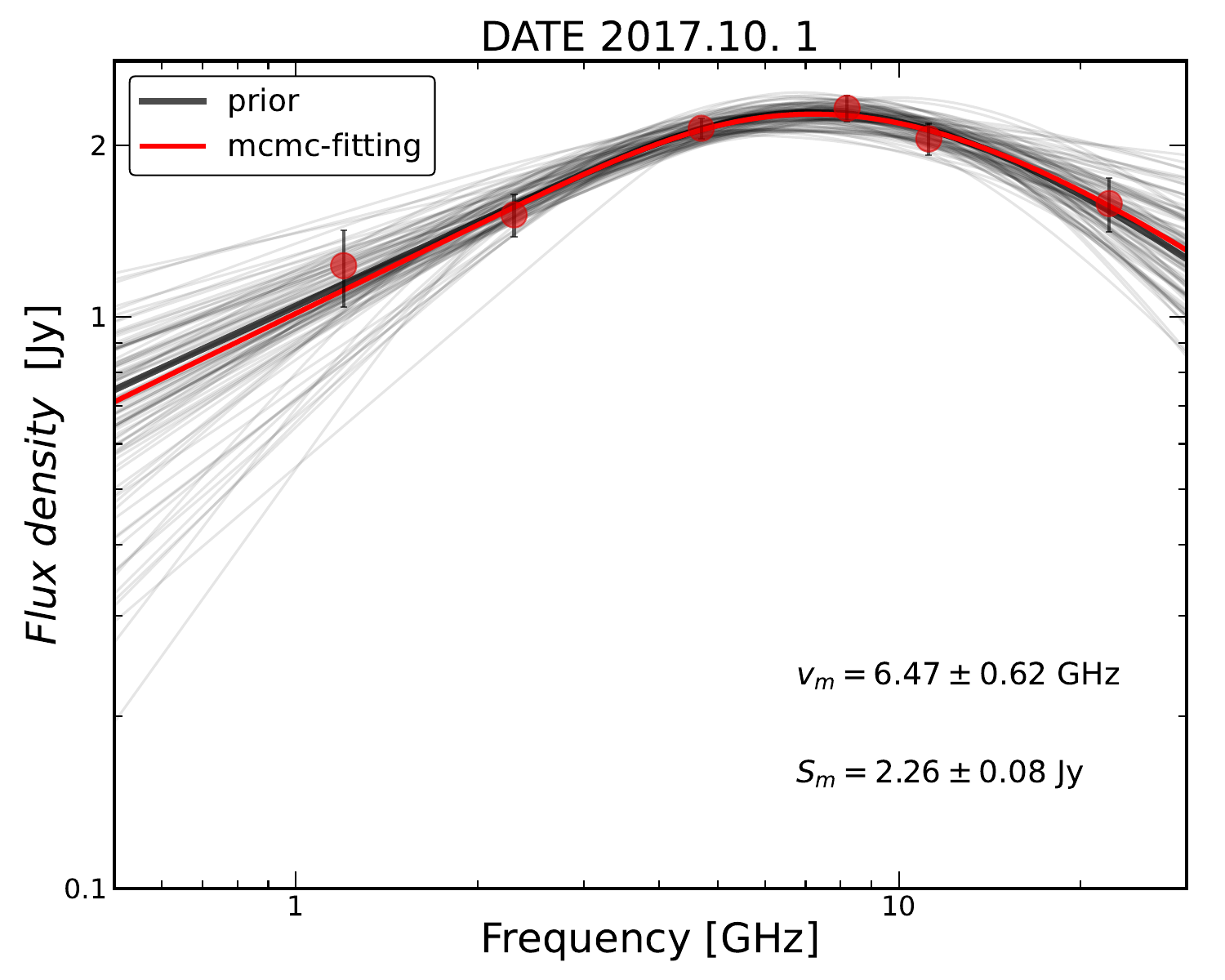}
\includegraphics[width=0.48\textwidth]{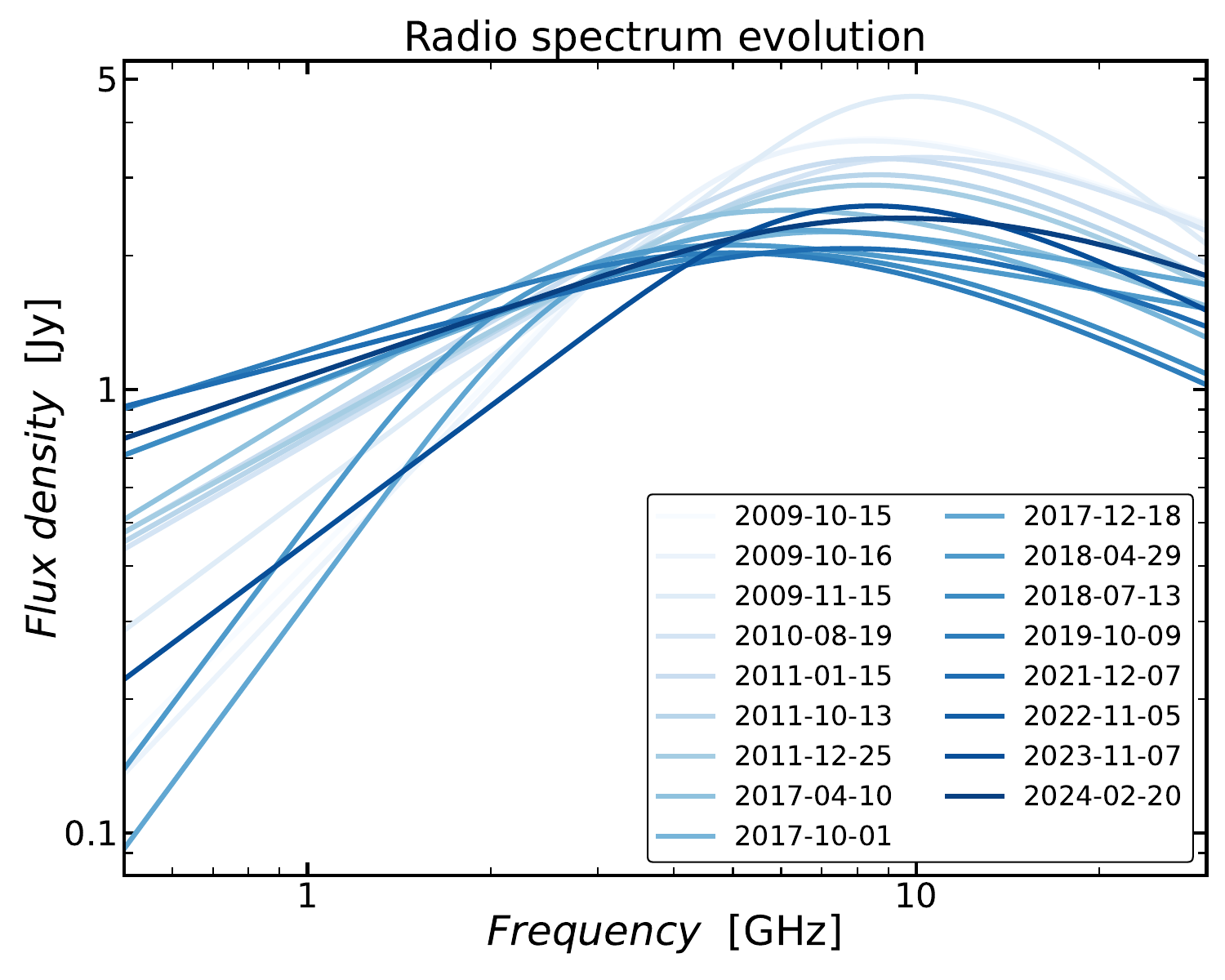}
\caption{Examples of Synchrotron Self-Absorption model fit (left panel) and the evolution of the fitted spectra (right panel). In the left panel, red, gray, and black lines denote the best mcmc-fitting result, 5000 sets of models randomly selected in the parameter space, and the prior parameters, respectively. The right panel collected the fitted results of 17 epochs from 2009 to 2024, with darker lines representing more recent observations. 
\label{fig:SSA}}
\end{figure*}

\begin{figure*}
\centering
\includegraphics[width=1\textwidth]{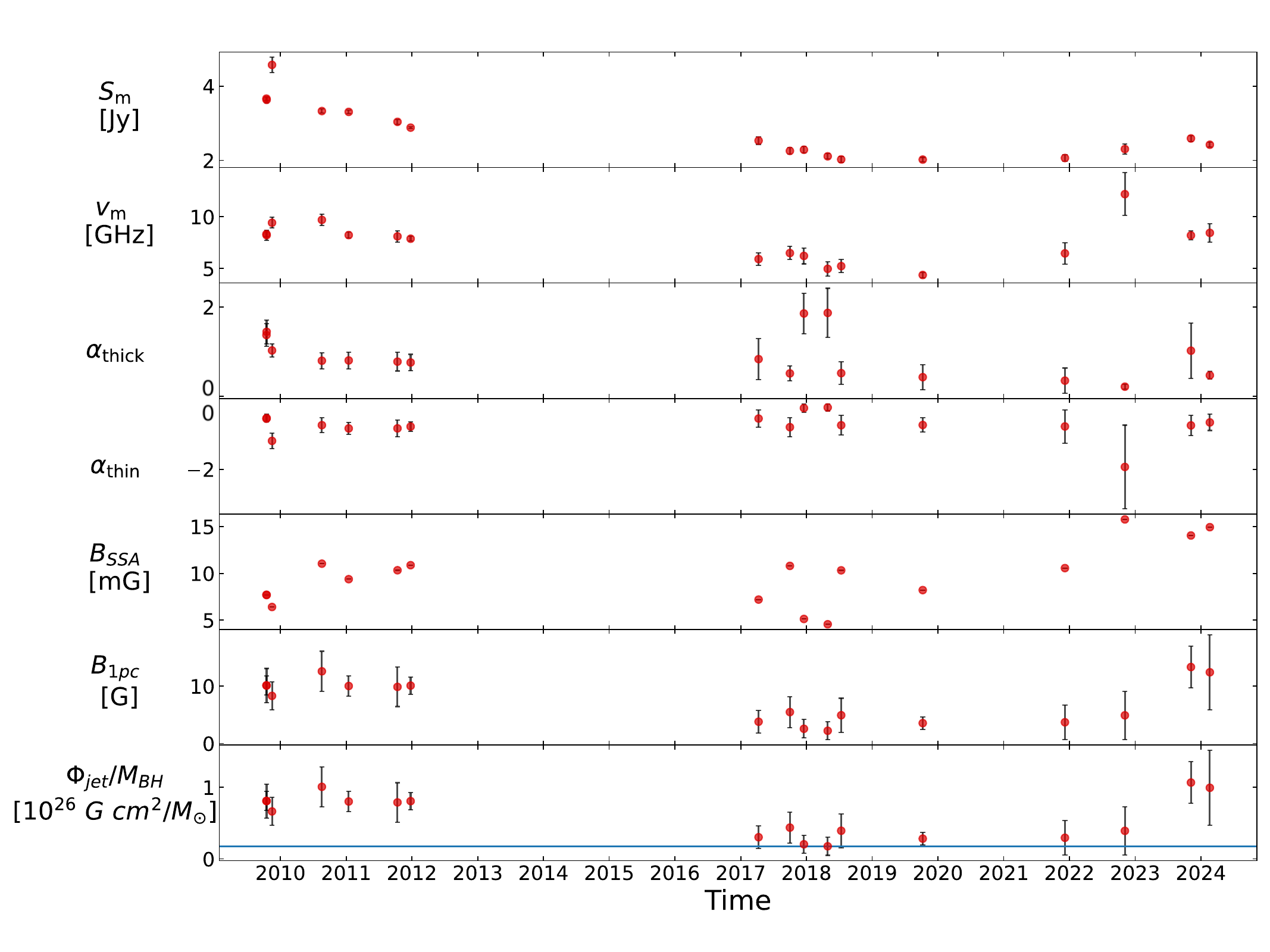}
\caption{Variations of the jet physical conditions with time. The upper four panels illustrate the time-dependent evolution of the synchrotron self-absorption spectrum parameters. The lower three panels display the calculated magnetic field strength ($B_{\rm SSA}$ and $B_{\rm 1pc}$) and magnetic flux, providing insights into the magnetic environment change over time. The horizontal line in the bottom panel denotes the predicted MAD magnetic flux. 
\label{fig-BandPhi}}
\end{figure*}

\begin{figure*}
\centering
\includegraphics[width=0.8\textwidth]{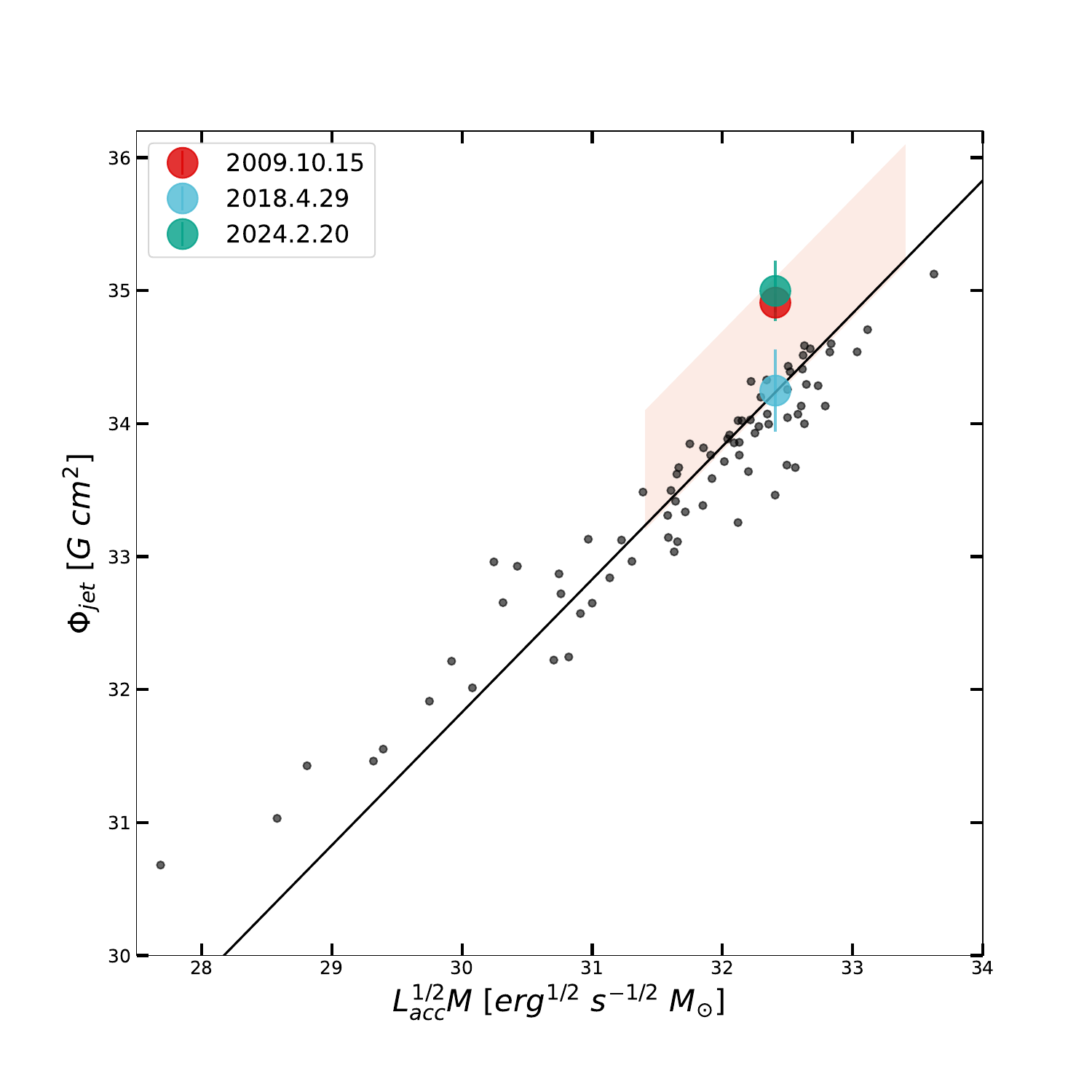}
\caption{Measured magnetic flux versus $L_{\rm acc}^{1/2} M_{\rm BH}$. The black circles are taken from \citet{2014Natur.510..126Z}, consisting of 76 radio-loud AGN. The parameter space for the OH~471 data is shown as a shaded area, and the data points on three representative epochs are shown as large circles.
\label{fig-relation}}
\end{figure*}

\begin{appendix} 

\section{RATAN-600 and RT-32 light curves}\label{app:ratan} 

The RATAN-600 radio telescope, a 600-meter ring-shaped multi-element antenna operating in transit mode \citep{1979S&T....57..324K,1993IAPM...35....7P}. This design allows for instantaneous broadband spectral observations from 1--22~GHz within 3--5 minutes when a source transits across the focal line where the receivers are located. The angular resolution, measured by the full width at half maximum (FWHM), depends on the antenna elevation angle, and the resolution along declination ${\rm FWHM}_{\rm Dec}$ is three to five times worse than that along the right ascension resolution ${\rm FWHM}_{\rm RA}$. The angular resolutions along RA and Dec calculated for the average angles are presented in Table~\ref{tab:ratan} for the six observing frequencies. Under favourable conditions and average elevation angles (${\rm Dec}\sim0^{\circ}$), the detection limit reaches around 5 mJy at 4.7 GHz for a RATAN-600 single sector for an integration time of $\sim3s$.

The measurements were processed using an automated data reduction system \citep{2016AstBu..71..496U,2018AstBu..73..494T} and the Flexible Astronomical Data Processing System (\textsc{FADPS}) standard package modules \citep{1997ASPC..125...46V} for the broadband RATAN-600 continuum radiometers. Flux density calibration relied on seven well-established secondary standards which are selected due to their high radio brightness and stable flux densities: 3C\,48 \citep{2010MNRAS.402...87A}, 3C\,147, 3C\,286 \citep{2017MNRAS.466..952A}, NGC\,7027, DR\,21, 3C\,295 and 3C\,309.1. The flux density scales were calculated based on \cite{1977A&A....61...99B} and \cite{2013ApJS..204...19P,2017ApJS..230....7P}. The measurements of the calibrators were corrected for the angular size and linear polarization based on the data from \citet{1994A&A...284..331O} and \citet{1980A&AS...39..379T}.

The total flux density uncertainty accounts for both the RATAN-600 calibration curve error, and the random antenna temperature measurement error \citep{2016AstBu..71..496U}, calculated as:
\begin{equation}
\left(\frac{\sigma_{\rm S}}{S_{\nu}}\right)^2 = \left(\frac{\sigma_{\rm c}}{g_{\nu}(h)}\right)^2 + \left(\frac{\sigma_{\rm m}}{T_{{\rm ant},\nu}}\right)^2,
\end{equation}
where $\sigma_{\rm S}$ is the total flux density standard error; $S_{\nu}$, the flux density at a frequency ${\nu}$; $\sigma_{\rm c}$, the standard calibration curve error, which is about 1--2\% and 2--5\% at 4.7 and 8.2 GHz respectively; $g_{\nu}(h)$, the elevation angle calibration function; $\sigma_{\rm m}$, the standard error of the antenna temperature measurement; and $T_{{\rm ant},\nu}$ is an antenna temperature. The systematic uncertainty of the absolute flux density scale (\mbox{3--10}\% at 1--22 GHz) is not included in the total flux error. 

\begin{table}
\caption{RATAN-600 continuum radiometer parameters: the central frequency $f_0$, the bandwidth $\Delta f_0$, the detection limit for point sources per transit $\Delta F$. ${\rm FWHM}_{\rm {RA} \times \rm {Dec}}$ is the angular resolution along RA and Dec calculated for the average angles.} 
\label{tab:ratan}
\centering
\begin{tabular}{cccr@{$\,\times\,$}l}
\hline
$f_{0}$ & $\Delta f_{0}$ & $\Delta F$ &  \multicolumn{2}{c}{FWHM$_{\rm {RA} \times \rm{Dec.}}$}\\
(GHz)    &   (GHz)           &  (mJy beam$^{-1}$)   &   \multicolumn{2}{c}{}  \\
\hline
 $22.3$ & $2.5$  &  $50$ & $0\farcm17$ & $1\farcm6$  \\ 
 $11.2$ & $1.4$  &  $15$ & $0\farcm34$ & $3\farcm2$ \\ 
 $8.2$  & $1.0$  &  $10$ & $0\farcm47$ & $4\farcm4$   \\ 
 $4.7$  & $0.6$  &  $8$  & $0\farcm81$ & $7\farcm6$   \\ 
 $2.25$  & $0.08$  &  $40$ & $1\farcm71$ & $15\farcm8$  \\ 
 $1.25$  & $0.08$ &  $200$ & $3\farcm07$ & $27\farcm2$ \\ 
\hline
\end{tabular}
\end{table} 

The RT-32 radio telescopes of the Institute of Applied Astronomy of the Russian Academy of Sciences (IAA RAS) at the Zelenchukskaya~(Zc) and Badary~(Bd) stations were used to measure the flux densities at frequencies of 5.05 and 8.63 GHz, with the bandwidth $\Delta f_{0}$=900~MHz for both receivers. 
The observations were conducted in the drift scan mode. One scan lasted $\sim$1 minute at 8.63 GHz and $\sim$1.5 minutes at 5.05 GHz with 1$s$ registration time, the FWHM=3\farcm9 and 7\farcm0 at 8.63 and 5.05 GHz respectively; flux density limit reaching of about 20~mJy per scan for both frequencies under optimal observation conditions. To increase sensitivity, the scanning cycle was repeated multiple times to form a continuous observing set. 

The observed data were processed using the original program package \textsc{CV} \citep{kharinov2012} and the Database of Radiometric Observations. Scans corrupted by weather or interference were filtered out. 
The remaining scans were averaged and fitted with a Gaussian curve after baseline subtraction using a parabolic approximation. 
The antenna temperature and its error were estimated from the Gaussian analysis of the averaged scan. The reference signal error of the noise generator is less than 1\% and is also included in the result. The flux density scales were calculated similarly to the RATAN-600 observations \citep{1977A&A....61...99B,2013ApJS..204...19P}. 3C\,48, 3C\,147, 3C\,295 and 3C\,309.1 were used as the reference sources. 

Figure \ref{fig-vlbi-and-ratan} bottom panel shows the RATAN-600 lightcurves of OH~471. Several data points observed by RT-32 at 8 and 5 GHz after September 2022 are added in the lightcurve. Data points that significantly diverge from the expected evolutionary trend, likely due to observational errors, are excluded from the analysis.

\section{VLBI Observations and results} \label{app:vlbi}

The visibility data have been calibrated using standard procedures in the PIMA \citep{2011AJ....142...35P} and AIPS software package \citep{2003ASSL..285..109G}. We only performed hybrid imaging interactively using self-calibration and deconvolution cycles in the DIFMAP software package \citep{1997ASPC..125...77S}. The final images were made using natural weighting after several rounds of phase-only self-calibration followed by 1-2 amplitude self-calibration runs. Data severely affected by Radio Frequency Interference (RFI), insufficient integration time or sparse (u, v) coverage, which could not yield high-quality images, were excluded from our analysis. The VLBI data analysis, as well as other data analyses in this research,  were used vlbi-pipeline  \footnote{\url{https://github.com/SHAO-SKA/vlbi-pipeline}} deployed on the computing platform of the China SKA Regional Centre \citep{2019NatAs...3.1030A, 2022SCPMA..6529501A}.

Following the imaging process, we fit circular Gaussian models to the self-calibrated visibility data for each epoch using the MODELFIT task in DIFMAP. This enabled a quantitative characterization of the milliarcsecond (mas)-scale core and jet components, including their flux densities, sizes, and relative positions over time. Elliptical or circular Gaussians were fitted to the self-calibrated visibility data. Typically 1--2 jet components in addition to the core are adequately used to model the visibility data at each epoch.  The fitted 8.4-GHz Gaussian parameters are summarized in Table \ref{tab:parameters}. The uncertainties of the fitted parameters were estimated following \citep{2020ApJ...896...63L}. By registering the jet component positions relative to the core, we measured proper motions to analyze jet kinematics. 

VLBI observations spanning decades have provided detailed insights into the parsec-scale structure of the blazar OH~471. Early 18 cm (1.67 GHz) observations revealed a compact 2.2 mas core along with hints of more extended emission that could account for the flux density offset between VLBI and single-dish measurements \citep{1980AJ.....85..668M}. Subsequent VLBI monitoring firmly established OH~471's characteristic core-jet morphology on milliarcsecond (mas) scales \citep{1992A&A...260...82G,1998AJ....115.1295K,2005AJ....130.2473K}, with extremely high brightness temperatures exceeding $10^{13}$ K inferred from 1.6 GHz space VLBI observations \citep{2015A&A...583A.100L}. High-frequency (15-43 GHz) ground-based and 1.6 GHz space VLBI images resolve the core into two components separated by 0.76 mas along a position angle of $\sim81\degr$ east \citep{2013AJ....146..120L,2015AJ....150...58F,2015A&A...583A.100L}.  While lower-frequency global and space VLBI observations detected an extended jet component stretching up to 30 mas \citep{1992A&A...260...82G, 2000A&A...364..391L}, higher-frequency VLBA images have failed to detect this distant emission, potentially due to resolution effects or a steepening spectral index.

\begin{figure*}
\includegraphics[width=0.98\textwidth]{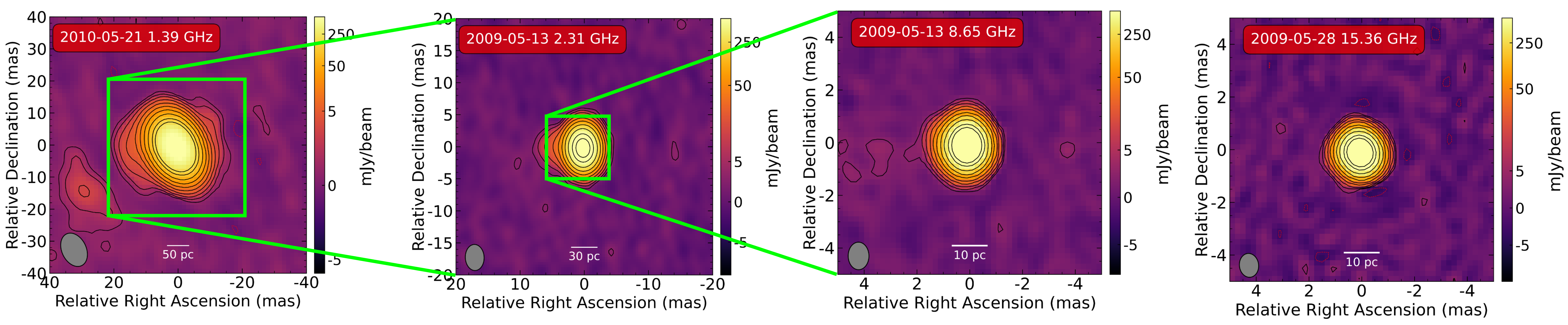}
\caption{Representative VLBI images of OH~471 across frequencies: This figure displays VLBI images at close epochs from right to left at frequencies of 15.4, 8.7, 2.3, and 1.4 GHz. Image details: 15.4 GHz with beam FWHM of 0.9 mas $\times$ 0.7 mas, PA = 6.9$^{\circ}$, a peak flux density of 3.04 Jy beam$^{-1}$ and rms noise of 0.98 mJy beam$^{-1}$; 8.7 GHz with beam FWHM of 1.0 mas $\times$ 0.8 mas, PA = 0.4$^{\circ}$, a peak flux density of 2.57 Jy beam$^{-1}$ and rms noise of 0.72 mJy beam$^{-1}$; 2.3 GHz with beam FWHM of 4.1 mas $\times$ 2.9 mas, PA=2.4$^{\circ}$, a peak flux density of 1.474 Jy beam$^{-1}$ and rms noise of 0.85 mJy beam$^{-1}$; and 1.4 GHz with beam FWHM of 10.9 mas $\times$ 7.6 mas, PA=25.6$^{\circ}$, a peak flux density of 0.83 Jy beam$^{-1}$ and rms noise of 0.15 mJy beam$^{-1}$. The lowest contour is set at three times the rms noise level, with contours increasing in a step of 2.
\label{fig:vlbiimages}}
\end{figure*}

Complementing these findings, the VLBI images presented in Figure \ref{fig:vlbiimages} provide finer scale views of the core-jet morphology within 2 mas across a range of frequencies, showing similar core-jet structures at different frequencies.
Figure \ref{fig-vlbi-and-ratan} shows the radio lightcurves derived from VLBI observations, which account for the total flux densities integrated over the entire core-jet region. 

\begin{figure*}
\centering
\includegraphics[width=0.8\textwidth]{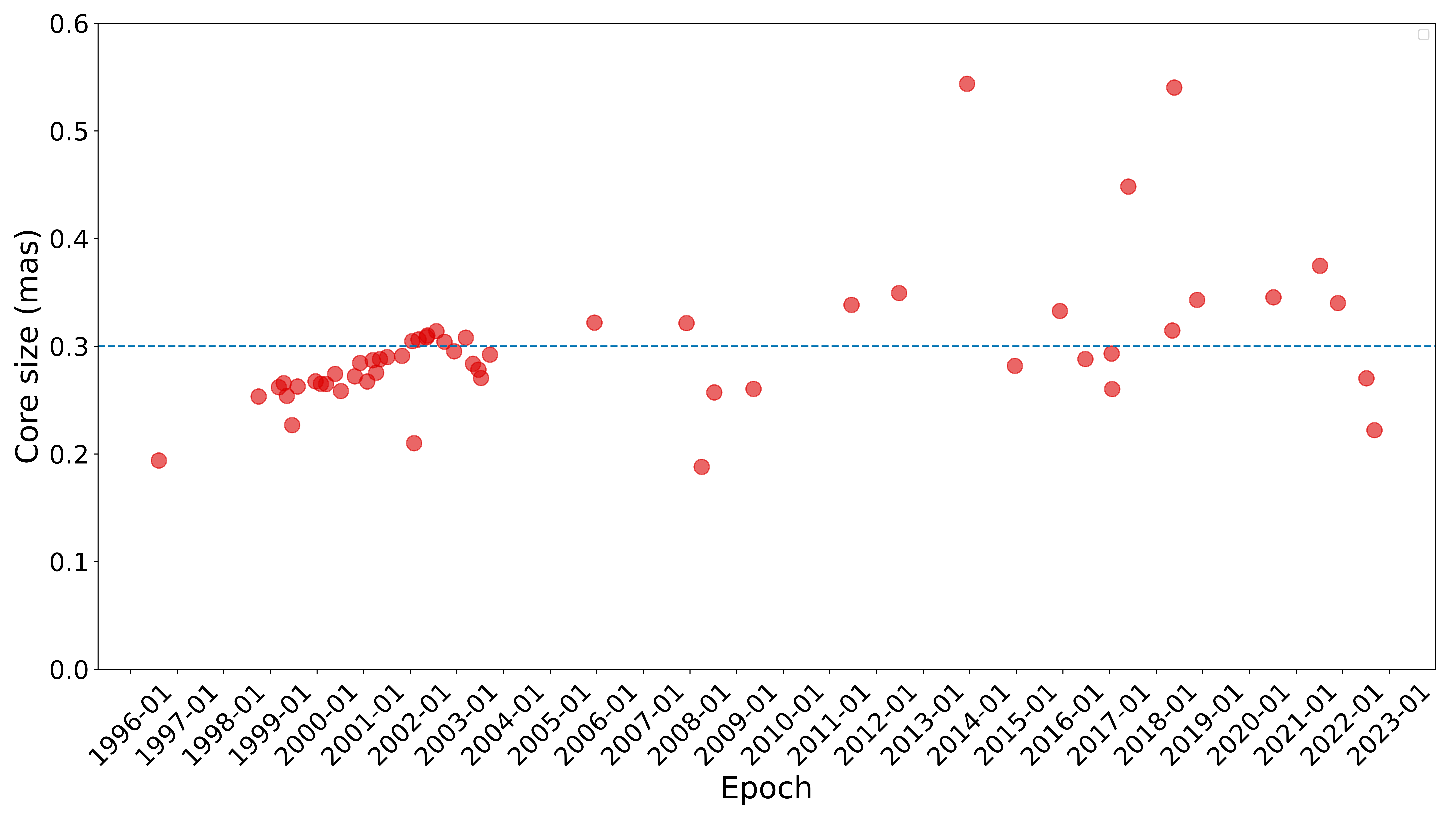}
\caption{Changes of core size of OH~471 over time. A dashed line represents the average core size ($0.30 \pm 0.11$ mas) from 8-GHz VLBI observations. Notably, three data points showing a core size greater than 0.4 mas are linked to observations with limited (u, v) coverage, where only the core, lacking any jet structure, was detected. This observational limitation influences the measured core sizes. 
\label{fig-coresize}}
\end{figure*}

\begin{table*}
    \centering
    \caption{Parameters of the fitted Gaussian model components }
    \begin{tabular}{ccccccc}  \hline
    Epoch  &  Freq & Comp  &  S     & R        & $\theta_{\rm FWHM}$  & $T_\mathrm{b}$ \\ 
           & (GHz) &       &(mJy)   &  (mas)   & (mas $\times$ mas)   & $(10^{11}K)$ \\ 
       (1) & (2)   &  (3)  &(4)     &  (5)     & (6)                & (7 ) \\\hline 
2008/04/02 & 8.65 & C & 2.74 $\pm$ 0.14 &      & 0.36 $\times$ 0.10 & 55.8 \\
           &      & J2 & 0.46 $\pm$ 0.02 & 0.36 & 0.46 $\times$ 0.46 &  \\
2008/07/09 & 8.65 & C & 2.93 $\pm$ 0.15 &      & 0.37 $\times$ 0.18 & 31.9 \\
           &      & J2 & 0.34 $\pm$ 0.02 & 0.42 & 0.51 $\times$ 0.51 &  \\
2009/05/13 & 8.65 & C & 2.76 $\pm$ 0.14 &      & 0.37 $\times$ 0.18 & 29.3 \\
           &      & J2 & 0.28 $\pm$ 0.01 & 0.47 & 0.55 $\times$ 0.55 &  \\
2012/06/27 & 8.64 & C & 2.22 $\pm$ 0.11 &      & 0.35 $\times$ 0.35 & 13.1 \\
           &      & J2 & 0.52 $\pm$ 0.03 & 0.62 & 0.43 $\times$ 0.43 &  \\
2013/12/11 & 8.64 & C & 1.86 $\pm$ 0.09 &      & 0.54 $\times$ 0.54 & 4.5 \\
           &      & J2 & 0.40 $\pm$ 0.02 & 0.62 & 0.67 $\times$ 0.67 &  \\
2014/12/20 & 8.67 & C & 1.64 $\pm$ 0.08 &      & 0.54 $\times$ 0.15 & 14.9 \\
           &      & J2 & 0.46 $\pm$ 0.02 & 0.84 & 0.43 $\times$ 0.43 &  \\
2015/12/08 & 7.62 & C & 1.45 $\pm$ 0.07 &      & 0.33 $\times$ 0.33 & 9.4 \\
           &      & J2 & 0.67 $\pm$ 0.03 & 0.82 & 0.41 $\times$ 0.41 &  \\
2016/06/25 & 7.62 & C & 1.37 $\pm$ 0.07 &      & 0.29 $\times$ 0.29 & 11.9 \\
           &      & J2 & 0.80 $\pm$ 0.04 & 0.78 & 0.51 $\times$ 0.51 &  \\
2017/01/16 & 8.65 & C & 1.42 $\pm$ 0.07 &      & 0.51 $\times$ 0.17 & 11.8 \\
           &      & J2 & 0.60 $\pm$ 0.03 & 0.84 & 0.55 $\times$ 0.55 &  \\
2017/01/21 & 8.67 & C & 1.43 $\pm$ 0.07 &      & 0.53 $\times$ 0.13 & 15.2 \\
           &      & J2 & 0.42 $\pm$ 0.02 & 0.90 & 0.35 $\times$ 0.35 &  \\
2017/05/27 & 8.67 & C & 1.41 $\pm$ 0.07 &      & 0.45 $\times$ 0.45 & 5.0 \\
           &      & J2 & 0.69 $\pm$ 0.03 & 0.83 & 0.57 $\times$ 0.57 &  \\
2018/05/07 & 8.65 & C & 1.34 $\pm$ 0.07 &      & 0.64 $\times$ 0.16 & 9.7 \\
           &      & J2 & 0.61 $\pm$ 0.03 & 0.87 & 0.58 $\times$ 0.58 &  \\
2018/05/22 & 7.62 & C & 1.18 $\pm$ 0.06 &      & 0.54 $\times$ 0.54 & 2.9 \\
           &      & J2 & 0.63 $\pm$ 0.03 & 0.83 & 0.54 $\times$ 0.54 &  \\
2018/11/18 & 8.65 & C & 1.09 $\pm$ 0.05 &      & 0.61 $\times$ 0.19 & 6.7 \\
           &      & J2 & 0.53 $\pm$ 0.03 & 0.88 & 0.56 $\times$ 0.56 &  \\
2020/07/07 & 8.64 & C & 0.91 $\pm$ 0.05 &      & 0.35 $\times$ 0.35 & 5.5 \\
           &      & J2 & 0.73 $\pm$ 0.04 & 0.88 & 0.56 $\times$ 0.56 &  \\
2021/07/07 & 8.64 & C & 0.92 $\pm$ 0.05 &      & 0.37 $\times$ 0.37 & 4.7 \\
           &      & J2 & 0.63 $\pm$ 0.03 & 0.95 & 0.68 $\times$ 0.68 &  \\
2021/11/24 & 8.64 & C & 1.11 $\pm$ 0.06 &      & 0.34 $\times$ 0.34 & 6.9 \\
           &      & J2 & 0.67 $\pm$ 0.03 & 0.95 & 0.69 $\times$ 0.69 &  \\
2022/07/05 & 8.64 & C & 1.04 $\pm$ 0.05 &      & 0.27 $\times$ 0.27 & 10.2 \\
           &      & J2 & 0.59 $\pm$ 0.03 & 0.97 & 0.67 $\times$ 0.67 &  \\
2022/09/06 & 8.64 & C & 1.09 $\pm$ 0.05 &      & 0.39 $\times$ 0.13 & 15.8 \\
           &      & J2 & 0.52 $\pm$ 0.03 & 1.00 & 0.60 $\times$ 0.60 &  \\
\hline
    \end{tabular}
    \label{tab:parameters}
\end{table*}

\begin{table*}[]
    \centering
    \begin{threeparttable}
    \caption{Parameters of the SSA fitting. }
\begin{tabular}{cccccccc}   \hline
Epoch  &  $S_\mathrm{m}$   & $\nu_\mathrm{m}$   & $\alpha_{ \rm thick}$ & $\alpha_{\rm thin}$ &  $B_{\rm SSA}$  & $B_{\rm 1pc}$ & $\Phi_{\rm jet}/M $  \\ 
  &   (Jy)   & (GHz)  &  &  &   (mG)  &  (G) & $ (10^{25} \text{G} \ \text{cm}^2 / M_{\odot})$  \\ 
            (1) & (2)  &  (3)      &(4)   &  (5)        & (6)        & (7 )  & (8)\\\hline 
2009/10/15 & 3.7 $\pm$ 0.1 & 8.3 $\pm$ 0.3 & 1.4 $\pm$ 0.3 & $-$0.6 $\pm$ 0.1 & 7.7 $\pm$ 1.2  & 10.1 $\pm$ 1.6 & 8.1 $\pm$ 1.3 \\
2009/10/16 & 3.6 $\pm$ 0.1 & 8.2 $\pm$ 0.5 & 1.4 $\pm$ 0.3 & $-$0.5 $\pm$ 0.1 & 7.7 $\pm$ 2.3  & 10.1 $\pm$ 3.0 & 8.1 $\pm$ 2.4 \\
2009/11/15 & 4.6 $\pm$ 0.2 & 9.4 $\pm$ 0.5 & 1.0 $\pm$ 0.1 & $-$1.2 $\pm$ 0.2 & 6.4 $\pm$ 1.9  & 8.3 $\pm$ 2.4 & 6.6 $\pm$ 1.9 \\
2010/08/19 & 3.3 $\pm$ 0.1 & 9.7 $\pm$ 0.5 & 0.8 $\pm$ 0.2 & $-$0.7 $\pm$ 0.2 & 11.0 $\pm$ 3.1  & 12.6 $\pm$ 3.5 & 10.1$\pm$ 2.8 \\
2011/01/15 & 3.3 $\pm$ 0.1 & 8.2 $\pm$ 0.3 & 0.8 $\pm$ 0.2 & $-$0.8 $\pm$ 0.2 & 9.4 $\pm$ 1.7  & 10.0 $\pm$ 1.8 & 8.0 $\pm$ 1.4 \\
2011/10/13 & 3.0 $\pm$ 0.1 & 8.1 $\pm$ 0.6 & 0.8 $\pm$ 0.2 & $-$0.8 $\pm$ 0.2 & 10.3 $\pm$ 3.6  & 9.9 $\pm$ 3.4 & 7.9 $\pm$ 2.7 \\
2011/12/25 & 2.9 $\pm$ 0.1 & 7.9 $\pm$ 0.2 & 0.8 $\pm$ 0.2 & $-$0.8 $\pm$ 0.1 & 10.9 $\pm$ 1.6  & 10.1 $\pm$ 1.5 & 8.1 $\pm$ 1.2 \\
2017/04/10 & 2.5 $\pm$ 0.1 & 5.9 $\pm$ 0.6 & 0.8 $\pm$ 0.5 & $-$0.6 $\pm$ 0.2 & 7.2 $\pm$ 3.8  & 3.8 $\pm$ 2.0 & 3.0 $\pm$ 1.6 \\
2017/10/01 & 2.3 $\pm$ 0.1 & 6.5 $\pm$ 0.6 & 0.5 $\pm$ 0.2 & $-$0.8 $\pm$ 0.3 & 10.8 $\pm$ 5.4  & 5.5 $\pm$ 2.7 & 4.4 $\pm$ 2.2 \\
2017/12/18 & 2.3 $\pm$ 0.1 & 6.2 $\pm$ 0.8 & 1.9 $\pm$ 0.5 & $-$0.3 $\pm$ 0.1 & 5.1 $\pm$ 3.2  & 2.6 $\pm$ 1.6 & 2.0 $\pm$ 1.3 \\
2018/04/29 & 2.1 $\pm$ 0.1 & 5.0 $\pm$ 0.7 & 1.9 $\pm$ 0.6 & $-$0.2 $\pm$ 0.1 & 4.6 $\pm$ 3.3  & 2.2 $\pm$ 1.6 & 1.8 $\pm$ 1.3 \\
2018/07/13 & 2.0 $\pm$ 0.1 & 5.2 $\pm$ 0.6 & 0.5 $\pm$ 0.2 & $-$0.7 $\pm$ 0.3 & 10.3 $\pm$ 6.2  & 4.9 $\pm$ 3.0 & 3.9 $\pm$ 2.4 \\
2019/10/09 & 2.0 $\pm$ 0.1 & 4.4 $\pm$ 0.3 & 0.4 $\pm$ 0.3 & $-$0.7 $\pm$ 0.2 & 8.2 $\pm$ 2.6  & 3.5 $\pm$ 1.1 & 2.8 $\pm$ 0.9 \\
2021/12/07 & 2.1 $\pm$ 0.1 & 6.5 $\pm$ 1.0 & 0.4 $\pm$ 0.3 & $-$0.8 $\pm$ 0.5 & 10.6 $\pm$ 8.6  & 3.7 $\pm$ 3.0 & 3.0 $\pm$ 2.4 \\
2022/11/05 & 2.3 $\pm$ 0.1 & 12.2 $\pm$ 2.1 & 0.2 $\pm$ 0.1 & $-$1.9 $\pm$ 1.2 & 15.8 $\pm$ 13.5 & 4.9 $\pm$ 4.2 & 3.9 $\pm$ 3.4 \\
2023/11/07 & 2.6 $\pm$ 0.1 & 8.2 $\pm$ 0.4 & 1.0 $\pm$ 0.6 & $-$0.8 $\pm$ 0.3 & 14.1 $\pm$ 3.9  & 13.3 $\pm$ 3.7 & 10.7 $\pm$ 2.9 \\
2024/02/20 & 2.4 $\pm$ 0.1 & 8.4 $\pm$ 0.9 & 0.5 $\pm$ 0.1 & $-$0.7 $\pm$ 0.2 & 14.9 $\pm$ 7.9  & 12.4 $\pm$ 6.5 & 9.9 $\pm$ 5.2 \\ 
\hline

\end{tabular}
\begin{tablenotes}
\item[Note] The uncertainties showed for fitting parameters including $S_\mathrm{m}$, $\nu_\mathrm{m}$, $\alpha_{ \rm thick}$, and $\alpha_{\rm thin}$ are fitting errors. The uncertainties of magnetic field feature measurements are transferred from the input parameters.
\end{tablenotes}

\end{threeparttable}
\label{tab:SSAandB}

\end{table*}

\section{Jet kinematics}

\begin{figure*}
\centering
\includegraphics[width=0.8\textwidth]{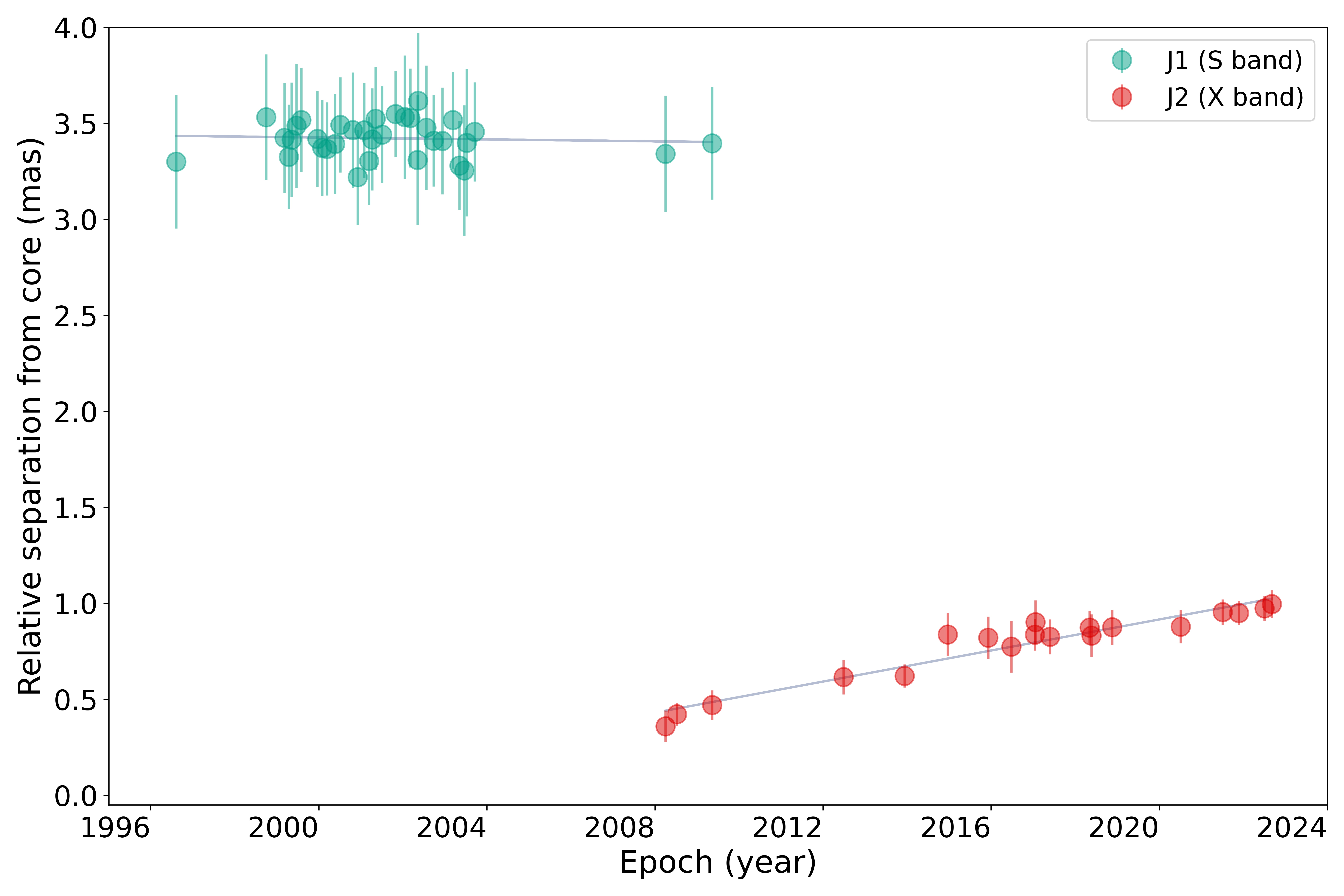}
\caption{Temporal changes in the positions of jet components J1 and J2 in OH~471. J1's positions are measured from the 2.3 GHz VLBI data, and J2's positions are measured from the 8.4 GHz VLBI data. Proper motions for J1 and J2 are represented through fitted trajectories: $\mu_{\text{J1}} = -0.0025\pm 0.0085 ~\text{mas yr}^{-1}$ (stationary), and $\mu_{\text{J2}} = 0.040  \pm 0.003  ~\text{mas yr}^{-1} (4.4\pm 0.4 \, c)$.}
\label{fig-propermotion}
\end{figure*}

Figure \ref{fig-propermotion} displays the proper motions of components J1 and J2. J1 does not show significant proper motion. In contrast, the inner jet J2 exhibits an outward motion of $0.040 \pm 0.003$ mas yr$^{-1}$ ($4.4 \pm 0.3 \,c$). As J1 was not detected proper motion based on 2.3 GHz VLBI data, their information is not presented in Table \ref{tab:parameters}.

To evaluate the relativistic beaming effect inside the target OH~471, we estimated the jet kinematics properties in different epochs. We determined the core brightness temperature using the equation (e.g., \citealt{Condon1982}):
\[
T_\text{b,obs} = 1.22 \times 10^{12} (1+z) \frac{S}{\nu^2 \theta_\mathrm{maj} \theta_\mathrm{min}} \rm K,
\]
where the flux density $S$ is in the unit of Jy, the observing frequency $\nu$ in  GHz, and $\theta$ is the elliptical/circular Gaussian component angular diameter (full-width half-maximum, FWHM) in mas. The brightness temperature results at 8 GHz are listed in Table \ref{tab:parameters}. From the measurement of brightness temperature, we could estimate the Doppler factor using 
$\delta=\frac{T_{\rm b,obs}}{T_{\rm b,int}}$ ,
where $T_{\rm b,int}$ represents the intrinsic brightness temperature. Assuming the energy equipartition between radiation particles and the magnetic field, \( T_{\rm b,int} \approx 5 \times 10^{10} \rm K \) \citep{1994ApJ...426...51R}. The corresponding Doppler factors are $25.0 \lesssim \delta \lesssim 49.9$.

The features of the jet, bulk Lorentz factor, $\Gamma $, and viewing angle to the line of sight, $\theta$, can be calculated from the following formulae \citep{1995PASP..107..803U}:
\[
\Gamma = \frac{\beta_a^2+\delta^2+1}{2\delta}, 
\]

\[
\rm tan{\theta} = \frac{2\beta_a}{\beta_a^2+\delta^2-1} .
\]
The $\beta_a$ in the formulae is jet proper motion speed. We measured from VLBI long-term monitoring at X-band and constrain this speed to be $\beta_a = 4.4 \pm 0.3c$. Adopting this result and the Doppler factor in different epochs, the bulk Lorentz factor and viewing angle are $12.4 \lesssim \Gamma \lesssim 25.2$ and $0.2\degr \lesssim \theta \lesssim 0.9 \degr$. These parameters are used in calculating the magnetic field strength for the corresponding RATAN observation epochs.

\section{Synchrotron self-absorption spectrum model fitting}\label{app:ssa}

The synchrotron self-absorption (SSA) spectrum is characterized by a distinctive shape: it rises steeply at low frequencies where the emission is self-absorbed, reaches a peak where the optical depth is approximately unity, and then falls off at higher frequencies where the emission becomes optically thin. The conventional SSA spectrum can be described by the following equation:
\begin{equation}
S(\nu) = S_{\text{m}} \left(\frac{\nu}{\nu_{\text{m}}}\right)^{2.5} \left[1-\exp\left(-\left(\frac{\nu}{\nu_{\text{m}}}\right)^{-\alpha - 2.5}\right)\right]
\label{eq:ssa}
\end{equation}
where $S_ \nu$ represents the flux density at frequency $\nu$; $S_{\text{m}}$ is the peak flux density at the turnover frequency; $\nu_{\text{m}}$ is the turnover frequency; $\alpha$ is the spectral index in the optically thin part of the spectrum.
The term $\left(\frac{\nu}{\nu_{\text{m}}}\right)^{2.5}$ describes the optically thick part of the spectrum where $\nu < \nu_{\text{m}}$, and the emission rises steeply with increasing frequency. The term in square brackets accounts for the transition from optically thick to thin emission, and the exponential term represents the attenuation of the emission due to self-absorption.
The modelled $\nu_\mathrm{m}$ constrains the electron density and magnetic field strength at the emission site. The peak flux $S_\mathrm{m}$ indicates total radiative power. The index  $\alpha_{\rm thin}$ provides clues about particle acceleration and cooling mechanisms.
The exact shape of the SSA spectrum can vary depending on the details of the electron energy distribution and the geometry of the emitting region. The spectral index $\alpha$ can also provide information about the energy distribution of the radiating electrons, which is often assumed to follow a power law.
When fitting an SSA spectrum to observational data, the values of  $S_{\text{m}}$, $\nu_{\text{m}}$, and $\alpha$ are typically free parameters that are adjusted to obtain the best fit to the data. This fit can then be used to infer the physical properties of the source, such as the magnetic field strength and electron density, as previously discussed.

\citet{2000A&A...361..850T} derived an SSA spectrum equation (their equation 41): 
\begin{equation}
S_{\nu} = S_\text{m} \left( \frac{\nu}{\nu_\text{m}} \right)^{\alpha_{\text{thick}}} \left[ \frac{1-\exp\left(-\tau_\text{m}\left(\frac{\nu}{\nu_\text{m}}\right)^{\alpha_{\text{thin}} - \alpha_{\text{thick}}}\right)}{1 - \exp(-\tau_\text{m})} \right]
\label{eq:turler}
\end{equation}
where, $\tau_\text{m} \approx \frac{3}{2} \left( 
\sqrt{1 - \frac{8 \alpha_{\text{thin}}}{3 \alpha_{\text{thick}}} }
-1 \right) $. 

The main difference between this equation \ref{eq:turler} and the conventional SSA spectrum equation \ref{eq:ssa} is the introduction of $\tau_\text{m}$ as an approximation of the optical depth at the turnover frequency $\nu_\mathrm{m}$, and as a function of the spectral indices $\alpha_{\text{thick}}$ and $\alpha_{\text{thin}}$. In the conventional SSA spectrum (Eq. \ref{eq:ssa}), the transition between the optically thick and thin regimes is often modelled with a simpler power-law dependence on frequency without explicitly accounting for the optical depth in this way. The equation \ref{eq:turler} represents a more sophisticated approach to modelling the SSA spectrum by accounting for the frequency dependence of optical depth, which can lead to a more accurate characterization of the physical conditions within the source, especially when the spectrum does not follow a simple power-law behaviour across the turnover point.  This change is made more reasonable for AGN by taking into account an inhomogeneous jet consisting of multiple homogeneous sub-regions, each with its own turnover frequency, resulting in a flatter optically thick spectral index than what would be expected from a homogeneous source. This inhomogeneity allows for a smoother transition between the optically thick and thin regimes of the spectrum. The assumptions of the model include a random orientation of the magnetic field within the jet, an electron energy distribution that follows a power law, and the application of general synchrotron radiation theory that allows the summing of emissions from various sub-regions. Moreover, relativistic effects such as Doppler boosting are assumed to be already included in the parameters, reflecting the complex motions within relativistic jets such as those from AGN. In addition, the use of the $\tau_\mathrm{m}$ approximation simplifies the determination of the optical depth at the turnover frequency. The inclusion of the turnover flux density $S_\mathrm{m}$ as a separate parameter allows for a direct fit to the observed peak of the spectrum, so that the model fitting accurately captures the peak of the SSA spectrum. 

In this work, we derive the SSA model parameters by applying the Bayesian Markov chain Monte Carlo (MCMC) algorithm with the $\texttt{Python}$ package $\texttt{emcee}$ \citep{2013PASP..125..306F}. The input prior parameter set is fitted from Linear Regression with the SSA model equation. Then we randomly select 5000 sets of parameters in the parameter space. The boundaries of the four parameter is: $0<S_\mathrm{m}<5$, $0<\nu_\mathrm{m}<20$, $0<\alpha_\mathrm{thick}<3$, and $-3<\alpha_\mathrm{thin}<0$. We then evaluate the fitting results with corner figures and mean square error (MSE). We select the spectrum with a critical peaked profile for the magnetic field measurement.

\section{SSA magnetic field strength}\label{app:mag}

Equation \ref{eq:turler} not only provides a phenomenological fit but can also be used to derive a direct physical interpretation. 
The magnetic field strength ($B$) can be estimated using the formula derived from the theory of synchrotron emission, which incorporates the turnover frequency ($\nu_{\rm m}$), the source size, and the assumption of equipartition of energy between the magnetic field and the particles \citep{1983ApJ...264..296M}:  
\begin{equation}
B_{\rm SSA} \approx 10^{-5} b(\alpha) \left(\frac{\nu_\mathrm{m}}{\text{GHz}}\right)^{5} \left(\frac{\theta}{\text{mas}}\right)^{4} \left(\frac{S_\mathrm{m}}{\text{Jy}}\right)^{-2} \left(\frac{\delta}{1+z}\right) 
\label{eq:Bssa}
\end{equation}
where $z$ is the redshift of the source; $B_{\rm SSA}$ represents the magnetic field strength in Gauss (G); $b(\alpha)$ is a dimensionless parameter depending on the optically thin spectral index and ranges from 1.8 to 3.8 based on the values given in \citet{1983ApJ...264..296M}; $\nu_\mathrm{m}$ is the turnover frequency in gigahertz (GHz); $\theta$ is the angular size of the emitting region in mas; 
$S_\mathrm{m}$ is the flux density at the turnover frequency in Jy. 
This equation assumes a specific geometry (e.g., a spherical emitting region) and that the magnetic field and particle energies are in equipartition.
$\nu_\mathrm{m}$ and $S_\mathrm{m}$ are obtained from SSA spectrum model fit (Appendix \ref{app:ssa}). $\theta$ and $\delta$ are from VLBI model fitting (Appendix \ref{app:vlbi}).

The uncertainty of the calculated $B_{\rm SSA}$ can be expressed as \(
\sigma_{B_{\rm SSA}} = f_{B_{\rm SSA}} \cdot B_{\rm SSA} \). Assessing $ f_{B_{\rm SSA}}$ can be made through error propagation: 
\[
f_{B_{\rm SSA}} =  \sqrt{\left(5 \cdot \frac{\sigma_{\nu_\mathrm{m}}}{\nu_\mathrm{m}}\right)^2 + \left(4 \cdot \frac{\sigma_{\theta}}{\theta}\right)^2 + \left(2 \cdot \frac{\sigma_{S_\mathrm{m}}}{S_\mathrm{m}}\right)^2}
\]

The magnetic field strength $B$ calculated from Equation \ref{eq:Bssa} is the value at the distance of the opaque core from the black hole. To provide some practical values at typical distances, one needs to convert a magnetic field strength measured at the location of the core (e.g., \(B_{\text{SSA}}\) s) to a value at 1 parsec from the black hole (\(B_\text{1pc}\)) usually involves understanding the scaling relationship of the magnetic field strength with distance in the jet.

The magnetic field in AGN jets often follows a power-law decay with distance from the central engine, expressed as:
\[ B(h) = B_0 \left( \frac{h}{h_0} \right)^{-\beta} \]
where:
\(B(h)\) is the magnetic field strength at distance \(h\) from the black hole,
\(B_0\) is the magnetic field strength at a reference distance \(h_0\) (often the core or a specific emission region),
\(\beta\) is the power-law index that describes how the magnetic field decreases with distance.

To convert \(B_{\text{SSA}}\) (measured at the core or the region associated with the spectrum turnover) to \(B_{1pc}\), one would need to know the distance of the core from the black hole in parsecs (or use the frequency to estimate this distance based on models of jet geometry) and apply the appropriate power-law index (\(\beta\)) that describes the magnetic field's radial profile in the jet.

The specific value of \(\beta\) can depend on the jet's physical conditions and the theoretical model used. For example, a commonly used model is the Blandford \& Königl model, which assumes a conical jet structure with a magnetic field decreasing as \(h^{-1}\) or \(h^{-2}\), depending on whether the magnetic field is dominated by the toroidal or poloidal component, respectively. In this study, $\beta=1$ is assumed.

The Doppler factor $\delta$ and redshift factor $(1+z)$ in equation \ref{eq:Bssa} play important roles in relating the observed properties to the intrinsic properties of the emitting region. The observed flux is boosted by $\delta^{3+\alpha}$ and observed frequencies by $\delta$, where $\alpha$ is the spectral index. So the Doppler factor needs to be included in the magnetic field strength equation to relate the observed quantities to the intrinsic ones. 
Given an observed brightness temperature \(T_{\text{obs}}\) and assumed \(T_{\text{int}}\), \(\delta\) can be calculated as \(\delta = T_{\text{obs}} / T_{\text{int}}\). Under the assumption of equipartition between particle and magnetic energy density, the upper limit of \(T_{\text{int}}\) can approximately to be $5 \times 10^{10}$ K \citep{1994ApJ...426...51R}. The $(1+z)$ factor accounts for converting between observed and rest-frame quantities.

Converting the angular size observed at a certain frequency using VLBI to that at the turnover frequency involves understanding the relationship between the observed frequency, the turnover frequency, and the angular size of the emitting region. The angular size of the emitting region at different frequencies can be related to the frequency by considering the optical depth (\(\tau\)). The angular size of the source is expected to decrease with increasing frequency in the optically thick regime due to the decreasing area of the emitting region that is above the optical depth threshold for significant emission.

A common approach to estimate the angular size at the turnover frequency (\(\theta(\nu_\mathrm{m})\)) from observations at another frequency (\(\theta(\nu)\)) involves the following equation:
\[ \theta(\nu_\mathrm{m}) = \theta(\nu) \left( \frac{\nu_\mathrm{m}}{\nu} \right)^{-k} \]
where \(\nu_\mathrm{m}\) is the turnover frequency, and \(\nu\) is the frequency of the VLBI observation, \(k\) is a constant that depends on the geometry of the emitting region and the spectral index. For a spherical, homogeneous synchrotron emitting source, \( k = 1/(1+\alpha) \) in the optically thick regime.

\section{magnetically arrested disk (MAD) }

\subsection{Foundational Principles of MAD Theory}

The Magnetically Arrested Disk (MAD) theory proposes that strong magnetic fields built up near the black hole can arrest inward accretion, dominate the dynamics of the inner disk, and launch energetic jets. As matter accretes, it drags along and amplifies the magnetic fields, leading to an accumulation of magnetic flux that exceeds the disk's diffusive capabilities. As a result, the magnetic pressure becomes comparable to the ram pressure of the gas, significantly influencing the accretion flow and forming a highly magnetized, arrested disk state. Strong MAD magnetic fields promote relativistic jet launching by extracting rotational energy from the black hole. MAD disks can also approach or exceed the Eddington luminosity limit, with magnetic fields playing a crucial role in collimating and driving very powerful jets. The process of magnetic flux accumulation allows episodic transitions to the MAD regime, characterized by the intermittent production of jet components.

\subsection{Theoretical Framework and Equations}

The calculation of the poloidal magnetic flux, \(\Phi_{\text{jet}}\), that threads a parsec-scale jet is crucial in understanding the dynamics of jets near accreting supermassive black holes. This calculation is foundational for assessing the magnetic fields' role in jet formation, acceleration, and collimation. 

The poloidal magnetic flux, \(\Phi_{\text{jet}}\), is a measure of the total magnetic field passing through a certain area perpendicular to the jet's flow direction. It is given by the integral of the magnetic field component (\(B_{\text{pol}}\)), which is aligned with the jet's length, over a cross-sectional area (\(A\)) of the jet:

\begin{equation}
\Phi_{\text{jet}} = \int_{A} B_{\text{pol}} \, dA
\label{eq:PhijetBpol}
\end{equation}

In the case of AGNs, it is often assumed that the magnetic field has a significant poloidal component that can be described by models of magnetohydrodynamic (MHD) flows. The strength of the magnetic field (\(B_{\text{pol}}\)) at different distances from the black hole is estimated using observations and theoretical models. This can involve complex modeling of the jet's emission properties, including Doppler boosting effects, and the physics of synchrotron radiation.  The cross-sectional area (\(A\)) through which the magnetic flux is calculated needs to be modelled or estimated based on observations. This often involves assumptions about the jet's geometry (e.g., cylindrical or conical shapes) and size at the parsec scale. 

VLBI observations can resolve the jet structure and sometimes directly measure the width of the jet at various points along its length.  If the jet's width can be directly resolved, this measurement can be used to estimate the cross-sectional area. The angular width of the jet, combined with the distance to the source, allows for a conversion to physical units (e.g., parsecs or light-years). For unresolved jets, the width might be estimated from the emission characteristics, such as the synchrotron self-absorption turnover, assuming a model for the jet's emission (e.g., a conical jet model) and the geometry of the emitting region.

The cross-sectional area $A$ depends on the assumed geometry of the jet. For a cylindrical jet, $A=\pi r^2$, where $r$ is the radius of the cylinder. For a conical jet, the area at a given distance from the central engine can be calculated if the opening angle of the cone and the distance along the jet axis are known, $A = \pi h^2 \tan^2 (\theta/2)$, where $h$ is the distance from the jet apex, $\theta$ is the observed angular size of the VLBI core. 

With estimates of \(B_{\text{pol}}\) and the jet's cross-sectional area, the poloidal magnetic flux can be calculated by integrating the magnetic field strength over this area.

The MAD condition can be quantitatively described by comparing the magnetic flux \(\Phi_{\text{BH}}\) threading the event horizon to the accretion rate \(\dot{M}\). A key dimensionless parameter that characterizes the MAD state is:
\begin{equation}
\phi = \frac{\Phi_{\text{BH} } }{\sqrt{\dot{M} r_\mathrm{g}^2 c}}
\label{eq:PhiMAD}
\end{equation}
where: \\
- \(\Phi_{\text{BH} }\) is the magnetic flux through the horizon, \\
- \(\dot{M}\) is the mass accretion rate, \\
- \(r_\mathrm{g} = GM/c^2\) is the gravitational radius of the black hole (with \(G\) being the gravitational constant, \(M\) the black hole mass, and \(c\) the speed of light).

In the MAD regime, the value of \(\phi\) becomes large, indicating a strong magnetic flux compared to the accretion rate. The exact threshold value for \(\phi\) indicating a transition to the MAD state can vary based on specific model assumptions. 
A typical value of $\phi = 50$ is often used \citep{2003PASJ...55L..69N, 2014Natur.510..126Z}.

While we cannot directly measure \(\Phi_{\text{BH} }\), we can observationally infer the poloidal magnetic flux threading parsec-scale jets, \(\Phi_{\text{jet} }\), by using the flux freezing approximation, i.e., \(\Phi_{\text{BH} } \approx \Phi_{\text{jet} } \).

\subsection{Jet magnetic flux and magnetic field strength}

The magnetic flux $\Phi_{\rm jet}$ is expressed in terms of the transverse-average magnetic field strength, $B$, as \citep{2015MNRAS.451..927Z}:
\begin{equation}
\begin{aligned}
    \Phi_{\text{jet}} &= \frac{2^{3/2} \pi r_\text{H} s h B (1 + \sigma)^{1/2}}{la} \\
    &= 8 \times 10^{33} f(a*) (1+\sigma)^{1/2} 
    \left[ \frac{M_\text{BH}}{10^9 M_\odot} \right]
    \left[ \frac{B_\text{1pc} } {1G} \right] 
\label{eq:Phijet}
\end{aligned}
\end{equation}

This equation involves several parameters: \\
- \(\Phi_{\text{jet}}\) is the jet magnetic flux. \\
- \(r_\text{H} = r_\mathrm{g} (1+(1-a^2)^{1/2}) \) represents the black hole event horizon radius. \\
- \(s\) is a parameter related to the angular frequency of the field lines compared to the black hole's angular frequency. \\
- \(h\) is the distance along the jet in parsecs. \\
- \(l\) is the black hole angular frequency \\
- \(a\) is the dimensionless black hole spin parameter.  The quantity \( f(a*) = r_\mathrm{H}/(a r_\mathrm{g}) \) is a function of the black hole spin.  For a rapidly rotating black hole, \( f(a*) \approx 1 \).  \\ 
- \(B\) is the magnetic field strength. \\
- \(\sigma = (\Gamma \theta_j / s)^2, s \lesssim 1\) is the jet magnetization parameter, representing the ratio of Poynting flux to kinetic energy flux. \( (\Gamma \theta_j) \sim 0.1 \), which led to similar $\sigma$ value with these in the literatures \citep[e.g.,][]{2021A&A...652A..14C, 2014Natur.510..126Z}.

To convert \(B_{\text{SSA}}\) to \(B\) at a specific location (e.g., 1 parsec from the black hole), one would use a model that describes how the magnetic field strength varies with distance within the jet. This could be a theoretical model based on jet physics or an empirical relationship derived from observations (see discussion in Appendix \ref{app:ssa}.
Once \(B_{1\text{pc}}\) is known, it can be used in the calculation of the jet magnetic flux using equation \ref{eq:Phijet}. The magnetic flux would typically be calculated based on the cross-sectional area of the jet at 1 parsec and the magnetic field strength across this area.

\end{appendix}

%
%

\bibliography{aa}
\bibliographystyle{aa}

\end{document}